\documentclass[10pt, conference, compsocconf]{IEEEtran}
\IEEEoverridecommandlockouts
\usepackage{cite}
\usepackage{amsmath,amssymb,amsfonts}
\usepackage{algorithmic}
\usepackage{graphicx}
\usepackage{textcomp}
\usepackage{xcolor}
\usepackage{subcaption}
\usepackage{multirow}
\usepackage{multicol}
\usepackage{float}
\usepackage{authblk}
\usepackage{url}

\def\BibTeX{{\rm B\kern-.05em{\sc i\kern-.025em b}\kern-.08em
    T\kern-.1667em\lower.7ex\hbox{E}\kern-.125emX}}
\usepackage[bookmarks=false]{hyperref}

\usepackage{listings}
\usepackage{DejaVuSansMono}
\usepackage[T1]{fontenc}

\ifCLASSINFOpdf
\else
\fi
\usepackage{tikz}	
\newcommand*\mycirc[1]{%
  \begin{tikzpicture}[baseline=(C.base)]
    \node[fill=black,text=white,draw,circle,inner sep=0.5pt](C) {#1};
  \end{tikzpicture}}

\newcommand{\workgroup}{octet}
\newcommand{\Workgroup}{Octet}
\newcommand{\workgroups}{octets}

\usepackage[absolute,showboxes]{textpos}

\TPMargin{5pt}
\newcommand{\copyrightstatement}{
    \begin{textblock}{0.84}(0.08,0.93)    
         \noindent
         \footnotesize
         \copyright 2019 IEEE. Personal use of this material is permitted. Permission from IEEE must be obtained for all other uses, in any current or future media, including reprinting/republishing this material for advertising or promotional purposes, creating new collective works, for resale or redistribution to servers or lists, or reuse of any copyrighted component of this work in other works. 
    \end{textblock}
}

\begin{document}  
\copyrightstatement
\title {Modeling Deep Learning Accelerator Enabled GPUs}
\author{\rm Md Aamir Raihan  \textsuperscript{*}}
\author{\rm Negar Goli \textsuperscript{*} \thanks{* equal contribution}}
\author{\rm Tor M. Aamodt}
\affil{Electrical and Computer Engineering\\University of British Columbia}
\affil{\{araihan, negargoli93, aamodt\}@ece.ubc.ca}


\maketitle

\begin{abstract}

The efficacy of deep learning has resulted in its use in a growing number of applications.
The Volta graphics processor unit (GPU) architecture from NVIDIA introduced a specialized
functional unit, the ``tensor core'', that helps meet the growing demand for higher performance for deep learning.
In this paper we study the design of the tensor cores in NVIDIA's Volta and Turing architectures.
We further propose an architectural model for the tensor cores in Volta.
When implemented a GPU simulator, GPGPU-Sim, our tensor core model
achieves 99.6\% correlation versus an NVIDIA Titan~V GPU in terms of
average instructions per cycle when running tensor core enabled GEMM workloads.  
We also describe support added to enable GPGPU-Sim to run CUTLASS, an
open-source CUDA C++ template library providing customizable GEMM templates that
utilize tensor cores. 
\end{abstract}

\section{Introduction}

Deep neural networks (DNNs) are having impact in a growing number of areas but
the benefits of DNNs come at the expense of high computational cost.  
Deep learning based data analytics has recently emerged as an
important technique\cite{Najafabadi2015}. 
DNNs have enabled breakthroughs in speech
recognition\cite{graves2013speech,bordes2012joint}, image recognition
\cite{ren2015faster,simonyan2014very}  and computer vision
\cite{vinyals2015show,kavukcuoglu2010learning}. 
DNNs require performing a large number of multi-dimensional matrix
(or tensor) computations.  Recent research has explored how to accelerate these operations 
\cite{jouppi2017datacenter,chen2014diannao,chen2016eyeriss,chakradhar2010dynamically,khan2008spinnaker,farabet2009cnp,kim2016neurocube,zhang2015optimizing}
and many companies are developing custom hardware for these workloads \cite{moloney2016embedded,zhang2016cambricon,tilley_2017}. 

GPUs are commonly used for deep learning, especially during training, as they
provide an order of magnitude higher performance versus a comparable investment
in CPUs~\cite{mlperf}.  
Specific effort has been directed at optimizing
GPU hardware and software for accelerating tensor operations found in DNNs.  On the hardware side, in the Volta architecture
NVIDIA introduced a specialized function unit called a Tensor Core for this purpose.  Tensor cores are also found on NVIDIA's more recent Turing
architecture \cite{Turing} 
and NVIDIA's T4 Turing-base GPUs are further optimized for inference tasks~\cite{nvidiaT4}.  
NVIDIA claims \cite{nvidiaTensorCores} tensor cores provide a speedup of $3\times$ on the Tesla V100 GPU when running mixed precision training.
Five out of six 2018 Gordon Bell Award Finalists employed tensor cores to improve application performance
and three did so specifically by accelerating machine learning \cite{gordonBell}.

However, to the best of our knowledge, the underlying design of tensor cores
has not been publicly described by NVIDIA.  Thus, we investigated the NVIDIA
tensor cores found in both Volta and Turing architectures.  Informed by our
analysis we extended GPGPU-Sim \cite{bakhoda2009analyzing} to include a model for tensor cores. 

This paper makes the following contributions:\\
\begin{itemize}
    \item It shows how different threads cooperate in transferring an input matrix to each tensor core.
    \item It gives an in-depth analysis of the execution of the tensor operation on the tensor cores and describes the microbenchmarks we used to perform our analysis. 
    \item It proposed a microarchitectural model of tensor cores consistent with the characteristics revealed through our microbenchmarks.  
    \item It describes our functional and timing model changes for modeling tensor cores in GPGPU-Sim.
    \item It describes support we added to enable applications built with NVIDIA's CUTLASS library to run on GPGPU-Sim.
    \item It quantifies the accuracy of our modified GPGPU-Sim by running tensor core enabled kernels generated with CUTLASS and thereby demonstrating an IPC correlation of 99.6\%.
\end{itemize}

We believe the observations made in this paper will provide useful guidance 
to those wishing to explore how to incorporate deep learning accelerators within GPUs.
The corresponding changes to model tensor cores in GPGPU-Sim should provide the academic community a helpful 
baseline for comparing alternative approaches.
The changes to enable CUTLASS to run on
GPGPU-Sim should ease study of architectural characteristics of custom
kernels on frameworks such as PyTorch (which was recently enabled to run on GPGPU-Sim
\cite{DBLP:journals/corr/abs-1811-08933}).


\section{Background}
This section briefly summarizes, at a high-level, relevant aspects of the Volta GPU architecture as documented by NVIDIA,
the source-code and instruction-level
interfaces for programming Tensor Cores on NVIDIA GPUs before finally describing 
what NVIDIA has disclosed about the design of their Tensor Cores.
\subsection{Volta Microarchitecture}
\begin{figure}
\centering
\centering\includegraphics[width=0.4\textwidth]{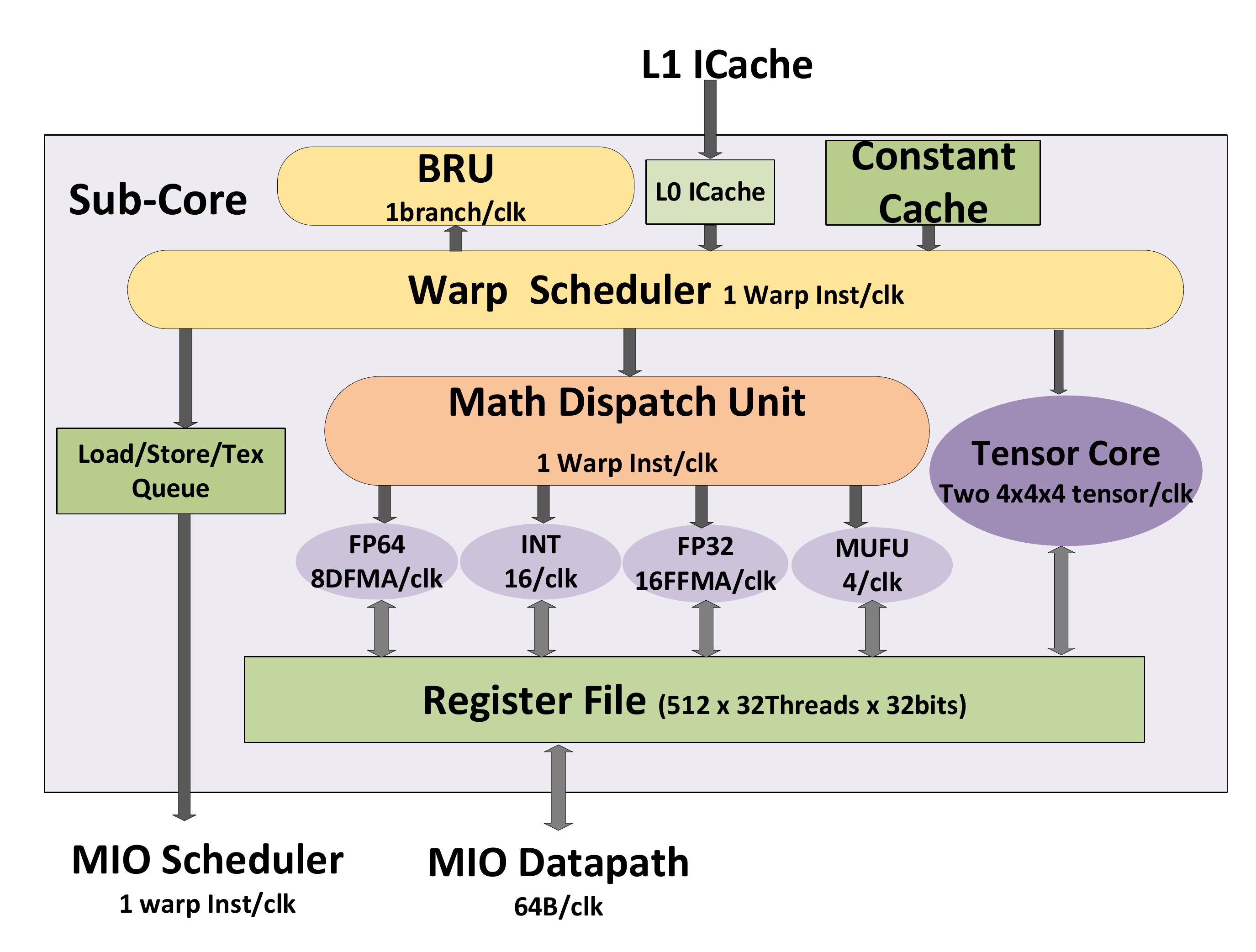}
\caption{Votla SM Sub-Core (reproduced from~\cite{VoltaProgrammability})}
\label{SubCoreMicroarchitecture}
\end{figure}

The first GPU to include accelerators for machine learning was NVIDIA's Volta~\cite{Volta}. 
Recent NVIDIA GPUs including Volta are generally composed of multiple Streaming Multiprocessors (SM) connected by 
an on-chip network to multiple memory partitions.  Each memory partition contains a portion of the 
last-level cache and connects the GPU to off-chip DRAM.
As described by NVIDIA, multiple tensor cores are included inside each SM.
The SM design in Volta is partitioned into four processing blocks which NVIDIA refers to as Sub-Cores. 
As shown in Figure~\ref{SubCoreMicroarchitecture}, each sub-core in Volta has two tensor cores, 
one Warp scheduler, one dispatch unit, and a 64~KB register file.

Besides the addition of tensor cores, Volta includes other enhancements relevant to performance of machine learning workloads:
In comparison to Pascal, NVIDIA's prior GPU architecture, each streaming multiprocessor (SM) in Volta has twice as many scheduling units along with
separate integer and 32-bit floating point (FP32) cores.  In addition, handling of divergent threads is different in Volta 
versus prior GPUs in that both paths following a branch can be executed by threads within a single warp in an interleaved 
fashion.  

NVIDIA typically releases several GPUs with the same underlying architecture but different amounts of on-chip resources.
For Volta, we focus in this paper on the Titan~V GPU.  The SM inside the Titan~V has the same number of registers as Pascal. 
However, the Titan~V GPU has 24 more SMs 
and thus can support more threads, warps, and thread blocks compared to prior generation GPUs. 

\vspace{0pt}

\subsection{Warp Matrix Function (WMMA) API}
\label{AA}

CUDA~9.0~\cite{nvidiadeveloperdocumentation} introduced a ``warp matrix function'' C++~language API to enable programmers to use the tensor cores on supported GPUs.  
This interface is also referred to as the CUDA C++ ``warp-level matrix multiply and accumulate''
(WMMA) API~\cite{wmma_api, nvidiaptxdeveloperdocumentation}.
It is well known that tiling can improve memory locality for dense matrix operations~\cite{Wolfe:1989:MIS:76263.76337}.
The WMMA~API exposes tensor cores to the GPU programmer as warp-wide operations for
performing the computation $D=A\times B+C$, where $A$, $B$, $C$ and $D$ can be tiles of larger matrices.
Using the WMMA~API, all threads in a warp cooperatively work together to perform 
a matrix-multiply and accumulate operation on these tiles.
NVIDIA's WMMA~API currently specifies a limited set of tile sizes. 
The sizes for tiles $A$, $B$, $C$ and $D$ are represented using the notation $M \times N \times K$, where $M \times K$ is the dimension of Tile~A, $K \times N$ is the dimension of Tile~B and thus $C$ and $D$ have dimension $M \times N$. 
CUDA~9.0 supports only one tile sizes, $16 \times 16 \times 16$,
 while later versions allow additional flexibility.

Using NVIDIA's terminology, each tile is further divided into ``fragments'' where a
fragment is a set of tile elements that are mapped into the registers of a single thread.
Thus, input matrices are distributed across different threads and
each thread contains only a portion of a tile. 
NVIDIA specifically states~\cite{nvidiadeveloperdocumentation} the mapping of tile elements to registers is unspecified.
Naively, considering a $16 \times 16$ tile contains 256 elements,
one possibility would be that each thread
in a warp with 32 threads would store an $\frac{256}{32}=8$ element fragment in eight separate general-purpose registers.
In Section~\ref{ReverseEngineeringTensorCore} we show that current GPU's do something more sophisticated.

The CUDA WMMA~API provides three new functions:
\lstinline{load_matrix_sync}, \lstinline{store_matrix_sync} and
\lstinline{mma_sync}. 
All three functions perform an implicit warp-wide barrier synchronization before computing a result.
The \lstinline{load_matrix_sync} and \lstinline{store_matrix_sync} functions are used for loading and storing a portion of the input matrices 
in the general-purpose registers accessible to each thread.  
The \lstinline{mma_sync} function performs a warp synchronous matrix multiply-accumulate operation producing an $M \times N$
(e.g., $16 \times 16$) result in the general-purpose registers associated with the tile for the D~matrix.

NVIDIA provides four high-level programming interfaces
for using tensor cores: the WMMA API described above and
three CUDA libraries: cuBLAS~\cite{cublas}, cuDNN~\cite{cuDNN,chetlur2014cudnn} 
and CUTLASS~\cite{cutlass,cutlasscode}.
In addition, many deep learning frameworks have included support for tensor cores \cite{abadi2016tensorflow, paszke2017pytorch}.  

\begin{figure*}[!t!]
\centering
\begin{minipage}{0.6\textwidth}
\begin{lstlisting}
    wmma.load.a.sync.layout.shape.type ra, [pa] {stride};
    wmma.load.b.sync.layout.shape.type rb, [pb] {stride};
    wmma.load.c.sync.layout.shape.type rc, [pc] {stride};
    wmma.mma.sync.alayout.blayout.shape.dtype.ctype rd, ra, rb, rc;
    wmma.store.d.sync.layout.shape.type rd, [pd] {stride};
\end{lstlisting}
\caption{Tensor Core PTX instructions}
\label{fig:ptx}
\end{minipage}
\end{figure*}

\subsection{PTX Instruction Set}
\label{sec:ptx}

NVIDIA's toolchain compiles CUDA into host code that runs on the CPU and device
code that runs on the GPU.  The device code is first compiled into a
device-independent machine-language instruction set architecture known as
Parallel Thread eXecution (PTX) before being compiled into device-specific 
machine code (SASS).

To perform operations on Tensor Cores at the
PTX level, NVIDIA introduced three PTX instructions in PTX version 6.0 \cite{nvidiaptxdeveloperdocumentation} with the syntax shown in
Figure~\ref{fig:ptx}.  
In this figure the ``\lstinline{sync}'' qualifier
indicates that the instruction waits for all threads in the warp to synchronize before
beginning execution.  
The PTX manual uses the term ``operand matrix'' to refer to a tile.
The ``\lstinline{layout}'' qualifier specifies whether an operand
matrix is stored in memory with a row-major or column-major layout.
The ``\lstinline{shape}'' qualifier represents the fragment size of the operand matrices (e.g., $16 \times 16 \times 16$ is specified
by setting \lstinline{shape} to \lstinline{m16n16k16}).
The ``\lstinline{type}'' qualifier indicates
the precision of the operand matrices, i.e. FP16 or FP32. 
For Volta, the A and B matrices must be FP16 but the C operand matrix can be either FP16 or FP32. 
NVIDIA's Turing architecture supports additional integer arithmetic modes initially targeted for inference.
In these, the operand matrices A and B can be 8, 4, or 1-bit signed or unsigned integers and operand matrices C and D are kept in higher-precision INT32 format to avoid overflow during accumulation~\cite{nvidiadeveloperdocumentation10}.

The operand matrices A, B and C must be loaded from memory to the register-file prior to
initiating a matrix-multiply operation.  This data movement is accomplished via three
\lstinline{wmma.load} PTX instructions.  Specifically, \lstinline{wmma.load.a},
\lstinline{wmma.load.b} and \lstinline{wmma.load.c} load the matrices A, B and C
respectively into registers \lstinline{ra}, \lstinline{rb} and \lstinline{rc} 
where \lstinline{ra}, \lstinline{rb} and \lstinline{rc} represent sets of general-purpose registers distributed across the threads of a warp
corresponding with the notion of a fragment.
\lstinline{pa}, \lstinline{pb}, \lstinline{pc} are the memory address where operand matrices A, B and C are stored in memory. 

Typically, input tiles loaded from memory are a portion of a larger matrix. 
To help accessing tiles of a larger matrix,
\lstinline{wmma.load} and \lstinline{wmma.store} support strided-memory access. 
The ``\lstinline{stride}'' operand specifies the beginning of each row (or column).


The \lstinline{wmma.mma} PTX instruction performs a warp-level matrix-multiply with accumulate
operation.  This instruction computes $D=A \times B+C$ using registers \lstinline{a}, \lstinline{b} and \lstinline{c} which
contain the matrix A, B and C respectively. The computed results are stored in
general-purpose registers \lstinline{d} in each thread. 

\begin{figure}
\centering
\centering\includegraphics[width=0.47\textwidth] {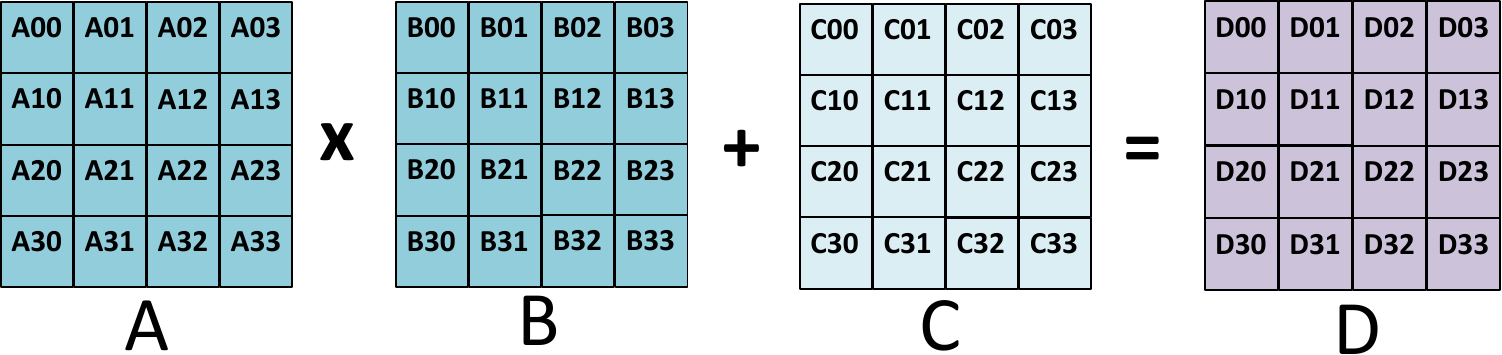}
	\caption{Tensor cores complete one $4\times4$ MACC operation per cycle ($D=A*B+C$).  Reproduces Figure~8 in\cite{Volta}.}
\label{TensorComputation}
\end{figure}

\subsection{Tensor Core}

Each tensor core is a programmable compute unit specialized for accelerating machine
learning workloads. The Tesla Titan V GPU contains 640 tensor cores distributed
across 80 SMs, with eight tensor cores per SM, providing a theoretical performance
of 125 TFLOPS at an operational frequency of 1530~MHz. 
According to NVIDIA \cite{Volta}, 
each tensor core can complete a single $4\times 4$ matrix-multiply-and-accumulation (MACC) 
each clock cycle, i.e. $D=A\times B+C$, where $A,B,C$ are $4\times 4$ matrices as shown in 
Figure~\ref{TensorComputation}. 
While individual tensor cores operate on $4\times 4$ matrices at any one time, 
as noted earlier, the WMMA API exposes the tensor cores on tile-sizes 
which are much larger.  Naively, a multiply of two $16 \times 16$ matrices
decomposes into a blocked matrix-multiply involving four $4 \times 4$ matrix-multiply accumulates
for each of the sixteen $4 \times 4$ submatrices of the result matrix.  Thus,
each \lstinline{mma_sync} at the CUDA C++ WMMA level or each \lstinline{wmma.mma} operation at the PTX level 
may be implemented with 64 separate tensor core operations.
The tensor cores have two modes of operation: FP16 and
mixed-precision. In FP16 mode, the tensor core reads three $4 \times 4$ 16-bit floating-point matrices as source operands
whereas in mixed-precision mode it reads two $4 \times 4$ 16-bit floating-point matrices along with a third
$4 \times 4$ 32-bit floating-point accumulation matrix.

\section{Demystifying NVIDIA's Tensor Cores}
\label{ReverseEngineeringTensorCore}

In this section we describe the results of our attempt to better understand the
low-level implementation details of tensor cores on recent GPUs.
Our analysis extends and refines that of Jia et al.~\cite{jia2018dissecting} who examined the distribution of matrix operand elements to registers 
for mixed precision mode in column-major layout. 
In their work, Jia et al.~\cite{jia2018dissecting} 
refer to a group of four consecutive threads within a warp as a ``thread group''.
We find it more convenient to shorten this to \textit{threadgroup}, which we do in the remainder of the paper. 
As there are 32 threads in a warp, there are $8$ \textit{threadgroups} in a warp.
We will refer to the \textit{threadgroup id}\footnote{Similar to ``group id'' in Jia et al.~\cite{jia2018dissecting}.}
of a given thread, which is given by $\lfloor\frac{threadIdx}{4}\rfloor$.  

\subsection{Microbenchmarks}\label{microbenchmarks}

In this section we discuss the microbenchmarks\footnote{\url{https://github.com/gpgpu-sim/gpgpu-sim_simulations/tree/master/benchmarks/src/cuda/tensorcore-microbenchmarks}} we used for analyzing the implementation of tensor cores.
We employ two types of microbenchmarks: Ones designed to determine how data move into and out of the tensor cores and others used
to determine how long the tensor cores take to perform operations.

\subsubsection{Fragment to thread mapping}

\begin{figure}[t]
\centering
\begin{minipage}{0.5\textwidth}
\begin{lstlisting}
    <FRAGMENT_DECLARATION> a_frag;
    wmma::load_matrix_sync(a_frag, mem_addr, stride );
    for(int i=0; i < a_frag.num_elements; i++) {
      float t=static_cast<float>(a_frag.x[i]);
      printf("THREAD%d CONTAINS %.2f\n",threadIdx.x,t);
    }
\end{lstlisting}
\caption{Microbenchmark for decoding thread fragments}
\label{fig:microbenchmark1}
\end{minipage}
\centering\includegraphics[width=0.32\textwidth] {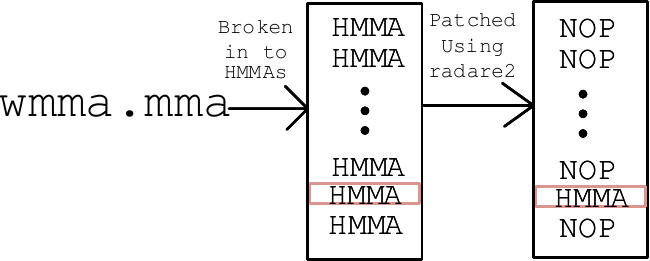}
\caption{Analyzing data accessed by tensor cores}
\label{DecodingSASS}
\centering\includegraphics[width=0.45\textwidth] {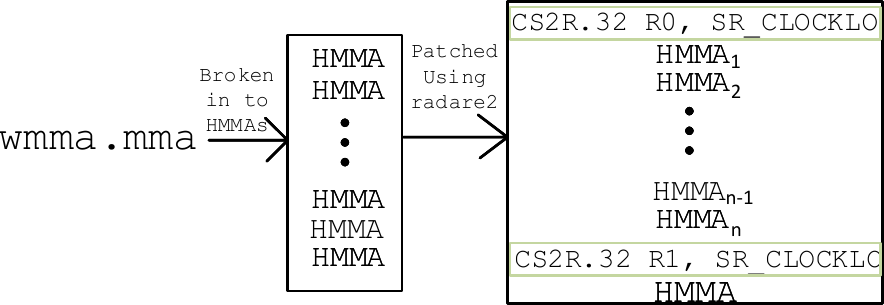}
\caption{Analyzing tensor core timing}
\label{TimingSASS}
\end{figure}

Figure~\ref{fig:microbenchmark1} contains a portion of the CUDA code employed
in Section~\ref{sec:matdist} to determine the mapping between operand matrix elements and
threads.  This code is part of a larger general matrix multiplication (GEMM)
kernel operating on a $16\times16$ matrices.  Each thread loads a segment of the input matrix 
and prints it to the output console. By initializing each element of the input
matrix with different values it is straightforward to uncover the mapping from 
operand matrix element to thread with a warp.

\subsubsection{Analyzing machine instructions} 

As described in detail in Section~\ref{sec:sass} \lstinline{wmma.mma} PTX instructions 
are mapped into multiple \lstinline{HMMA} SASS instructions.  
Figure~\ref{DecodingSASS} illustrates, at a high level, the operation of our microbenchmarks
used for analyzing how data is accessed by \lstinline{HMMA} instructions.
We use radare2~\cite{radare2}
to replace all \lstinline{HMMA} operations except one with ``no operation'' (NOP)
instructions. 
Figure~\ref{TimingSASS} illustrates at a high-level the approach used by 
our microbenchmarks for analyzing the timing of low level operations on tensor cores.  
To develop these microbenchmarks we used radare2 to 
add code that reads the clock register before the $1^{st}$ and after the
$n^{th}$ HMMA instruction.

\subsection{Operand matrix element mapping}
\label{sec:matdist}

In this section we summarize the results of our
analysis of the distribution of matrix elements to threads.

\begin{figure*}[tb]
\centering
\begin{subfigure}{0.45\textwidth}
\centering\includegraphics[width=1\textwidth]{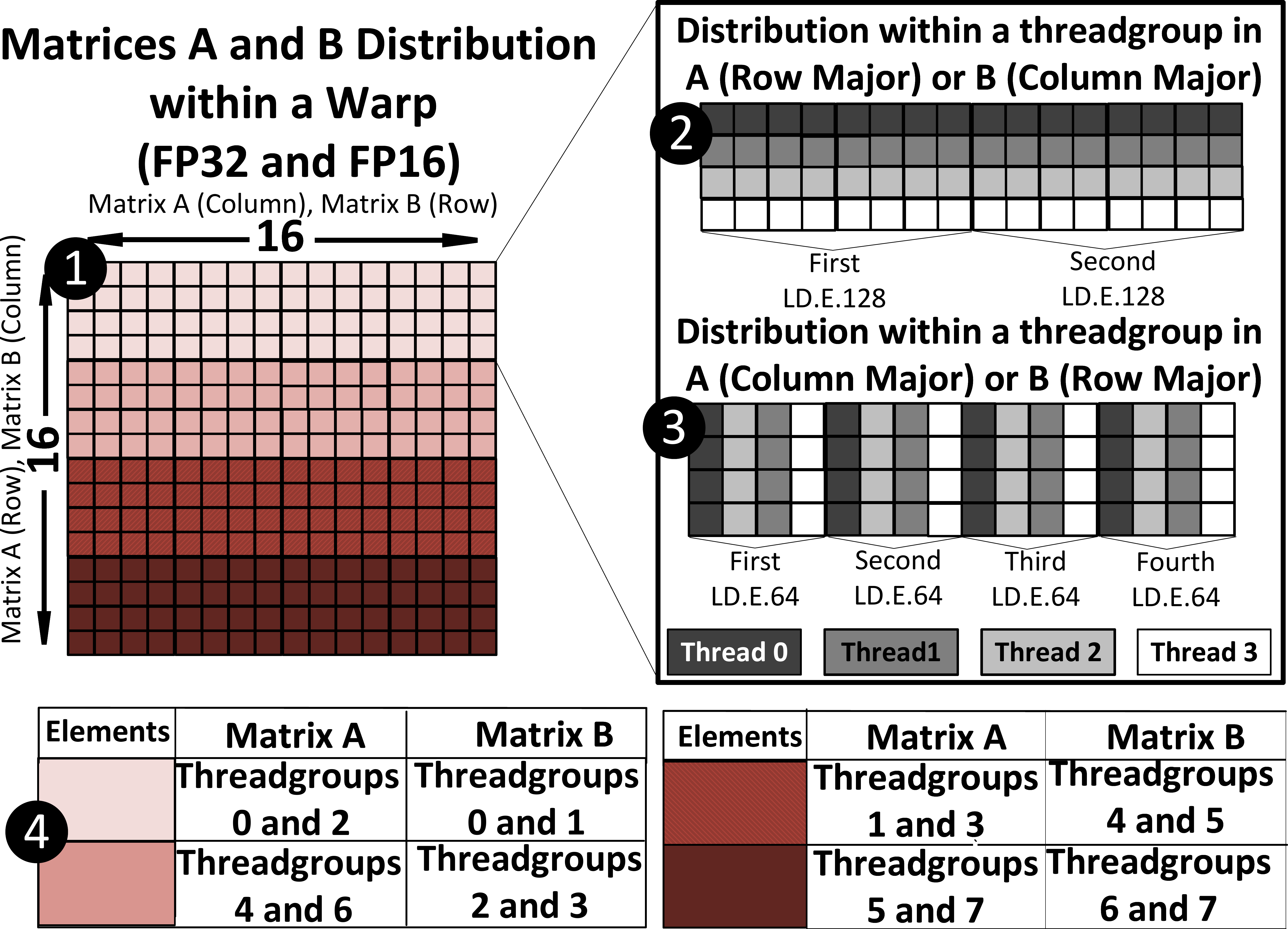}
\caption{Operand matrices A and B.}
\label{Mapping:a}
\end{subfigure}\vspace{0pt}
\hspace{0.1in}
\begin{subfigure}{0.45\textwidth}
\centering
\centering\includegraphics[width=1\textwidth]{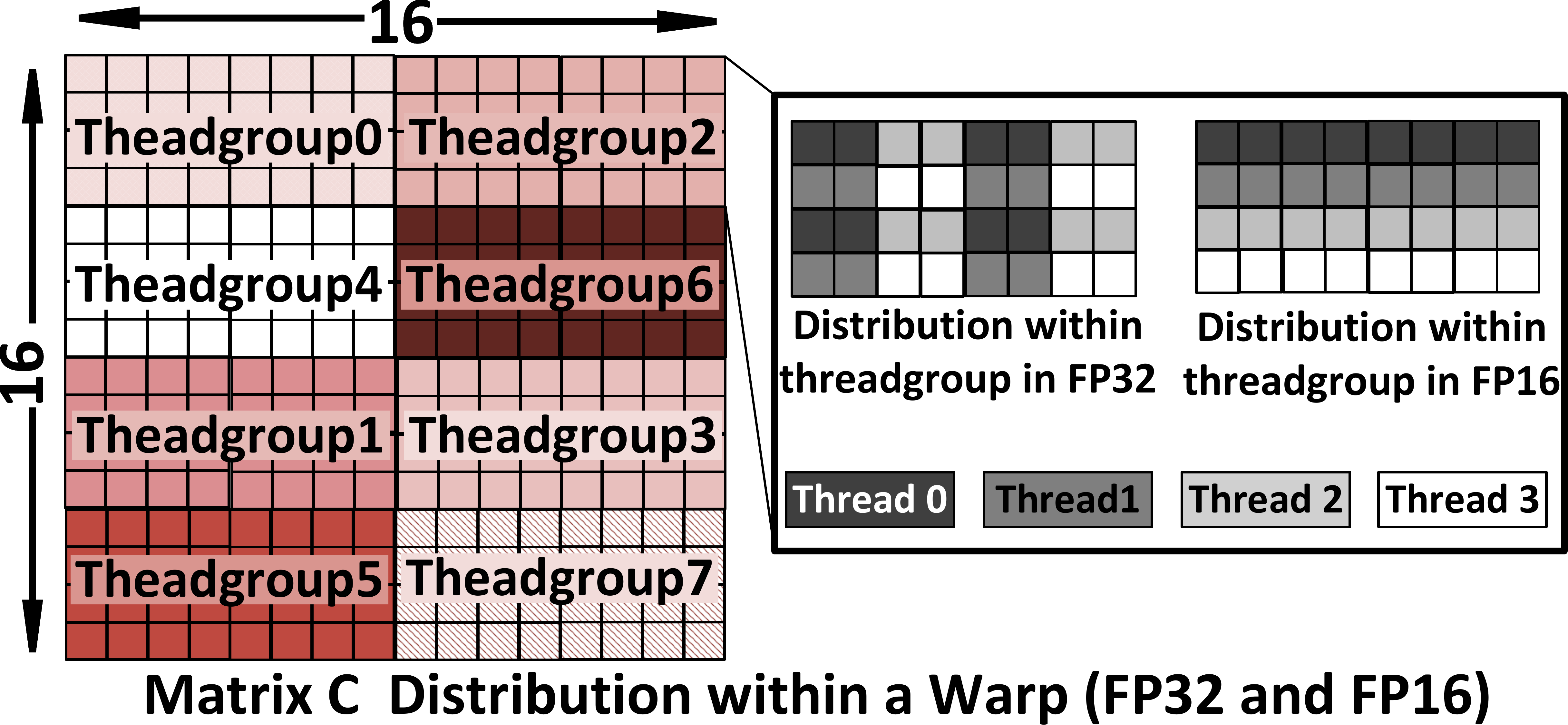}
\caption{Operand matrix C.}
\label{Mapping:b}
\end{subfigure}\vspace{0pt}
\caption{Distribution of operand matrix elements to threads for Tensor Cores in the Titan V (Volta).}
\label{Mapping}
\end{figure*}

\subsubsection{Volta Tensor Cores}

Figures~\ref{Mapping:a} and \ref{Mapping:b}
summarize how the elements of matrix operands are mapped to the registers of 
individual threads within a warp.
The large rectangle (\mycirc{\small{1}})
represents $16\times16$ operand matrix A or B for both FP16 and mixed-precision modes of operation.
Smaller squares are individual elements of the operand matrix and elements in the same row 
are stored contiguously in memory.  
Each \textit{threadgroup} loads a different $4\times16$ sub-matrix, 
which we will refer to as a segment.
The four segments that make up the operand matrix are highlighted with different shading.

The upper right-hand portion of Figure~\ref{Mapping:a} (\mycirc{\small{2}},\mycirc{\small{3}}) shows how the elements
within a segment are distributed among the threads of a threadgroup.
Our analysis found that on Volta, each segment is loaded by two
different \textit{threadgroups}. Thus, each element of the A and B operand matrices 
are loaded by two different threads in a warp on Volta.
The bottom portion of Figure~\ref{Mapping:a} (\mycirc{\small{4}}) combined with the top-left portion (\mycirc{\small{1}})
summarize the exact mapping. For example, we found the first four consecutive rows of 
operand matrix A are loaded by \textit{threadgroup} 0 and 2.  

The distribution of matrix elements to threads for operand matrix A stored in row-major layout is the same as the
distribution of operand matrix B stored in column-major layout and vice-versa. 
For the operand matrix A in row-major layout, each thread inside the \textit{threadgroup} loads
$16$ consecutive elements using two coalesced 128-bit wide load instructions (\mycirc{\small{2}}) 
whereas in column
major layout each thread inside the \textit{threadgroup} loads four blocks of
four consecutive elements via four coalesced 64-bit wide load instructions, each
with a stride distance of 64 elements (\mycirc{\small{3}}).  

As illustrated in Figure~\ref{Mapping:b},
the distribution of matrix elements to threads is different for operand matrix C.
Specifically, for operand matrix C each \textit{threadgroup} loads a
$8\times 4$ segment of the matrix C.  Also, the specific distribution within the
threadgroup now depends on whether the matrix C is FP16 or FP32 and is independent of the layout.  
32-bit wide (partially coalesced) load instructions are used to access elements of matrix C in both modes of operation.

\subsubsection{Turing Tensor Cores}

\begin{figure*}[t!]
\centering\includegraphics[width=0.8\textwidth]{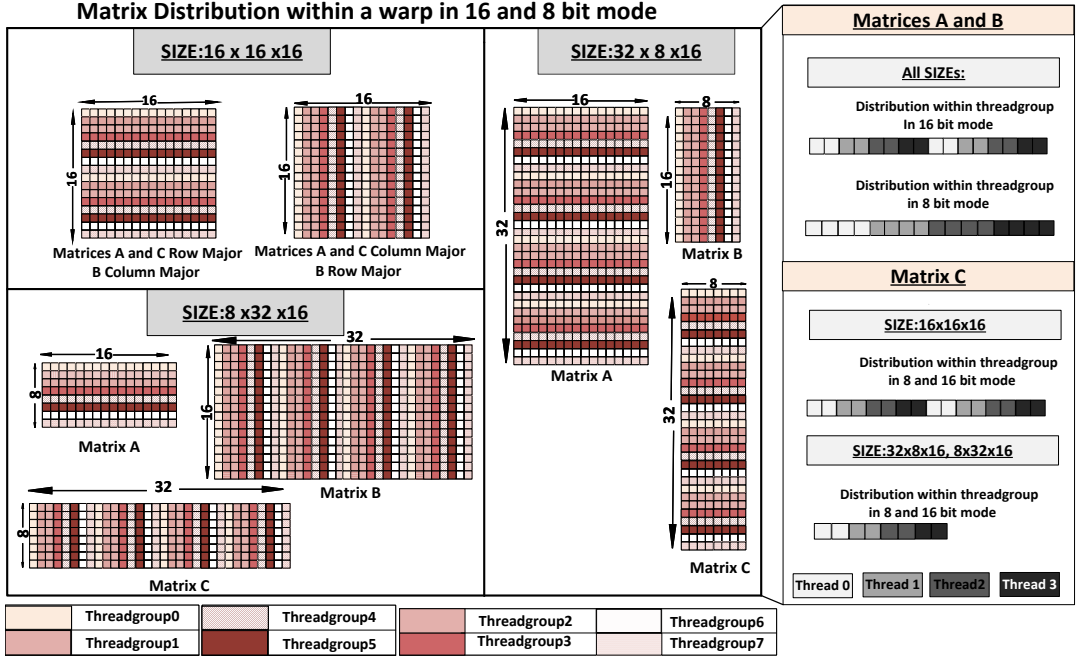}
\caption{Distribution of operand matrix elements to threads for tensor cores in the RTX 2080 (Turing).}
\label{Turing_mapping}
\end{figure*}

Figure \ref{Turing_mapping} summarizes the distribution of operand matrix
elements to threads for tensor cores in NVIDIA's Turing architecture.
Turing's tensor cores support three new precision modes: 1-bit, 4-bit and 8-bit,
along with three new tile sizes: $32\times 8\times 16$ and $8\times 32\times 16$ for 8
and 16-bit modes and $8\times 8\times 32$ for 4-bit mode.
Support for 1-bit mode was only enabled very recently as of this writing and
did not appear to work on our system. Thus, no analysis is provide for 1-bit mode in the rest of this paper.
We found Turing has a simpler distribution of elements to threads than Volta.
Specifically, each operand matrix element is loaded only once.  
Both tile size $32\times 8\times 16$ and
$8\times 32\times 16$ employ the same distribution.
For all modes and configurations, each row or column
(depending on the mode and operand matrix) is loaded by a \textit{threadgroup}
and consecutive \textit{threadgoups} load consecutive rows or columns.

\subsection{Machine ISA interface}
\label{sec:sass}

This section summarizes what we learned about how Tensor Cores are accessed
at the machine instruction set architecture level.  This level is typically called SASS for NVIDIA GPUs.  
The analysis here is based upon examining SASS disassembly using NVIDIA's \lstinline{cuobjdump} tool.

We found that \lstinline{wmma.load} and \lstinline{wmma.store} PTX instructions
are implemented by being broken into a group of normal SASS load (\lstinline{LD.E.64}, 
\lstinline{LD.E.128}, 
\lstinline{LD.E.SYS}) and 
store (\lstinline{ST.E.SYS}) instructions.  This suggests that Tensor Cores
access operand matrix fragments directly from the normal GPU register file.
In more detail, we found the \lstinline{wmma.load.c} PTX instruction is broken
into a group of \lstinline{LD.E.SYS} instructions.
For operand matrices A and B, depending on whether the operand matrix layout is row major or column major,
\lstinline{wmma.load} PTX instructions are broken into either four 64-bit loads~(\lstinline{LD.E.64}) or two 
128-bit loads (\lstinline{LD.E.128}), respectively.

\begin{figure}
\centering
\begin{subfigure}{.5\textwidth}
\centering\includegraphics[width=\textwidth]{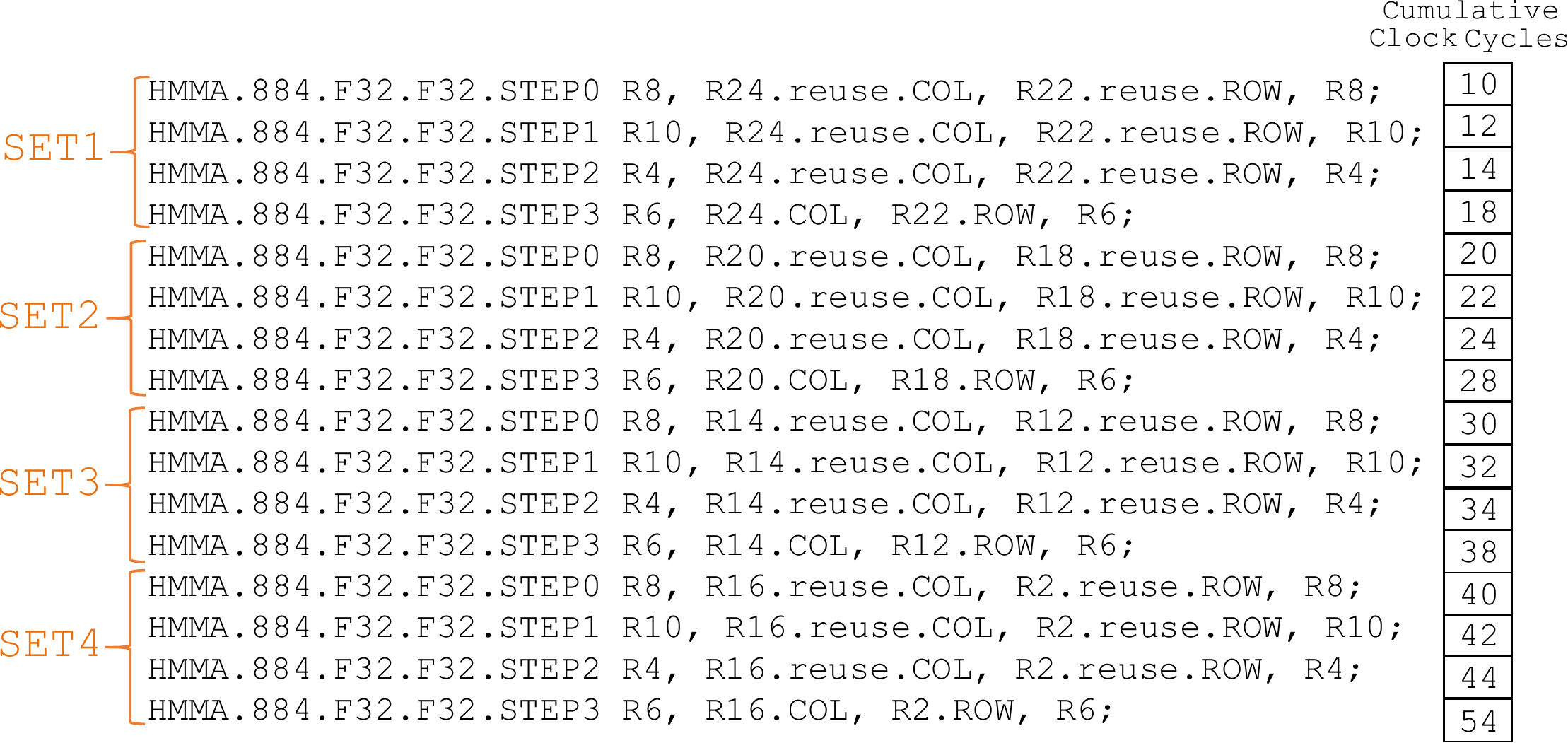}
\caption{Disassembled SASS instructions for Mixed precision mode}
\label{DisassembledSASS_32}
\end{subfigure}\vspace{0.15in}
\begin{subfigure}{.5\textwidth}
\centering\includegraphics[width=\textwidth]{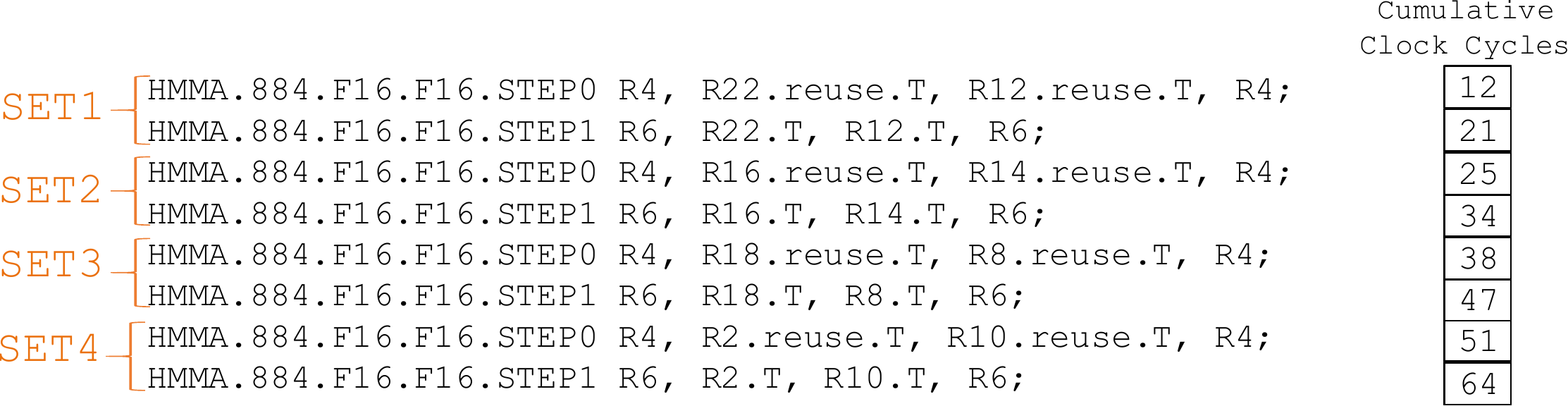}
\caption{Disassembled SASS instructions for FP16 mode}
\label{DisassembledSASS_16}
\end{subfigure}\vspace{0.15in}
\caption{Disassembled SASS instructions corresponding to WMMA:MMA API}
\label{DisassembledSass}
\end{figure}

Figure~\ref{DisassembledSass} illustrates the SASS code for Volta corresponding
with a single \lstinline{wmma.mma} PTX instruction. As can be seen in this figure,
matrix-multiply accumulate operations are implemented 
via a new SASS instruction, \lstinline{HMMA}.
Each \lstinline{HMMA} instruction has four operands and each operand uses a pair of registers.
By comparing the registers used by the \lstinline{HMMA} and the loads and stores,
we have inferred that a pair of adjacent registers accessed by different memory operations
are encoded in the \lstinline{HMMA} instruction using a single register identifier.
For example, \lstinline{R8} in the first \lstinline{HMMA} instruction in Figure~\ref{DisassembledSass}
appears from our analysis to represent the register pair \lstinline{<R8,R7>}.
The higher register identifier in the register pair is
the one encoded in the instruction.  For example, for the \lstinline{HMMA} instruction on the first line of
of Figure~\ref{DisassembledSass}, the destination register \lstinline{R8} actually
represents the pair \lstinline{<R8,R7>}.  Similarly, the remaining register identifiers actually represent three
pairs of source operand registers (\lstinline{<R24,R23>}, \lstinline{<R22,R21>} and \lstinline{<R8, R7>}).
Each of the four pairs of registers corresponds to operand matrices A, through D.

Some registers are annotated with ``\lstinline{reuse}'' in Figure~\ref{DisassembledSass}.
Gray~\cite{sgemm_implementation} analyzed NVIDIA's SASS instruction set for the earlier Maxwell architecture where
a similar annotation often appears.  
Based upon his analysis and related papers from NVIDIA on register file caching for GPUs~\cite{gebhart2011compile},
we believe the ``\lstinline{reuse}'' notation indicates 
the associated operand is reused in the
next step and therefore cached in the operand reuse cache to avoid a
register fetch and possibly to reduce bank conflicts.

\subsubsection{Volta Tensor Cores}
Each \lstinline{wmma.mma} PTX instruction is broken into a group of
\lstinline{HMMA} instructions. 

Figure~\ref{DisassembledSASS_32} illustrates the SASS code for
mixed precision mode.  In this mode, each PTX \lstinline{wmma.mma} instruction
is broken into 16 \lstinline{HMMA} instructions.  These are organized as four
{\em sets} of four \lstinline{HMMA} instructions.  Each \lstinline{HMMA}
instruction is annotated with ``\lstinline{STEP<n>}'' where \lstinline{<n>}
ranges from 0 to 3.  
Thus, each set comprises four steps.  
Figure~\ref{DisassembledSASS_16} illustrates the SASS code for FP16 mode in which
a single PTX \lstinline{wmma.mma} instruction is broken into four sets
consisting of only 2 steps.
Figure~\ref{DisassembledSass} also shows the cumulative clock cycles for the Volta
Tensor Cores. The latency of \lstinline{wmma.mma} API in mixed precision mode is
ten cycles lower than in FP16 mode.

\subsubsection{Turing Tensor Cores}
For Turing, each PTX \lstinline{wmma.mma} instruction is broken into a group of four HMMA instructions
for all modes except 4-bit where it is converted into a single HMMA instruction.
Table~\ref{table:3} shows the cumulative clock cycles for HMMA
instructions on the Turing architecture.
For $16\times 16\times 16$ tile size, the latency of \lstinline{wmma.mma} in mixed precision mode on Turing, 99 cycles (Table~\ref{table:3}),
is more than on Volta, 54 cycles (Figure~\ref{DisassembledSASS_32}).
The latency of mixed precision mode is
more than FP16 mode. 8-bit mode is fastest. The latency of 4-bit
mode is the highest, which may be because it is an experimental feature on the 2080~RTX.

\begin{table}[h]
 \centering
\begin{tabular}{ |c|c|c|c|c|c|} 
 \hline
  \multirow{2}{*}{TILE SIZE}&\multirow{2}{*}{PRECISION} & \multicolumn{4}{c|}{Average Cumulative Clock Cycles}  \\  \cline{3-6}
  \rule{0pt}{10pt}(MxNxK) &  & SET 1 & SET 2 & SET 3 & SET 4 \\
  \hline
  \multirow{3}{*}{16x16x16} & 16Bit (FP32 Acc)  & 42& 56& 78& 99\\ 
        & 16Bit (FP16 Acc) & 44& 52& 60& 74\\
       &  8Bit & 40& 44& 47& 59\\    
  \hline 
    \multirow{3}{*}{32x8x16} & 16Bit (FP32 Acc) & 48& 60& 81& 104\\
		& 16Bit (FP16 Acc) & 44& 52& 60& 74\\	
		& 8Bit & 52& 55& 59& 73\\
 \hline		
	\multirow{3}{*}{8x32x16}& 16Bit (FP32 Acc) & 42& 56& 77& 99\\	
		& 16Bit (FP16 Acc) & 42& 50& 58& 72\\	
		& 8Bit & 38& 42& 46& 56\\
 \hline		
	\multirow{1}{*}{8x8x32}& 4Bit & 230& - & - & -\\		
 \hline 
\end{tabular}
\caption{Average cycles to execute all HMMA instructions up to SET $n$ on Turing. ``Acc'' is accumulation mode.}\vspace{0pt}
\label{table:3}
\end{table}\vspace{0pt}

\subsection{HMMA Instruction Analysis}
This section explores HMMA execution in greater detail.

\subsubsection{Volta}

\begin{figure}[t]
\centering
\begin{subfigure}{.3\textwidth}
\centering\includegraphics[width=1\textwidth]{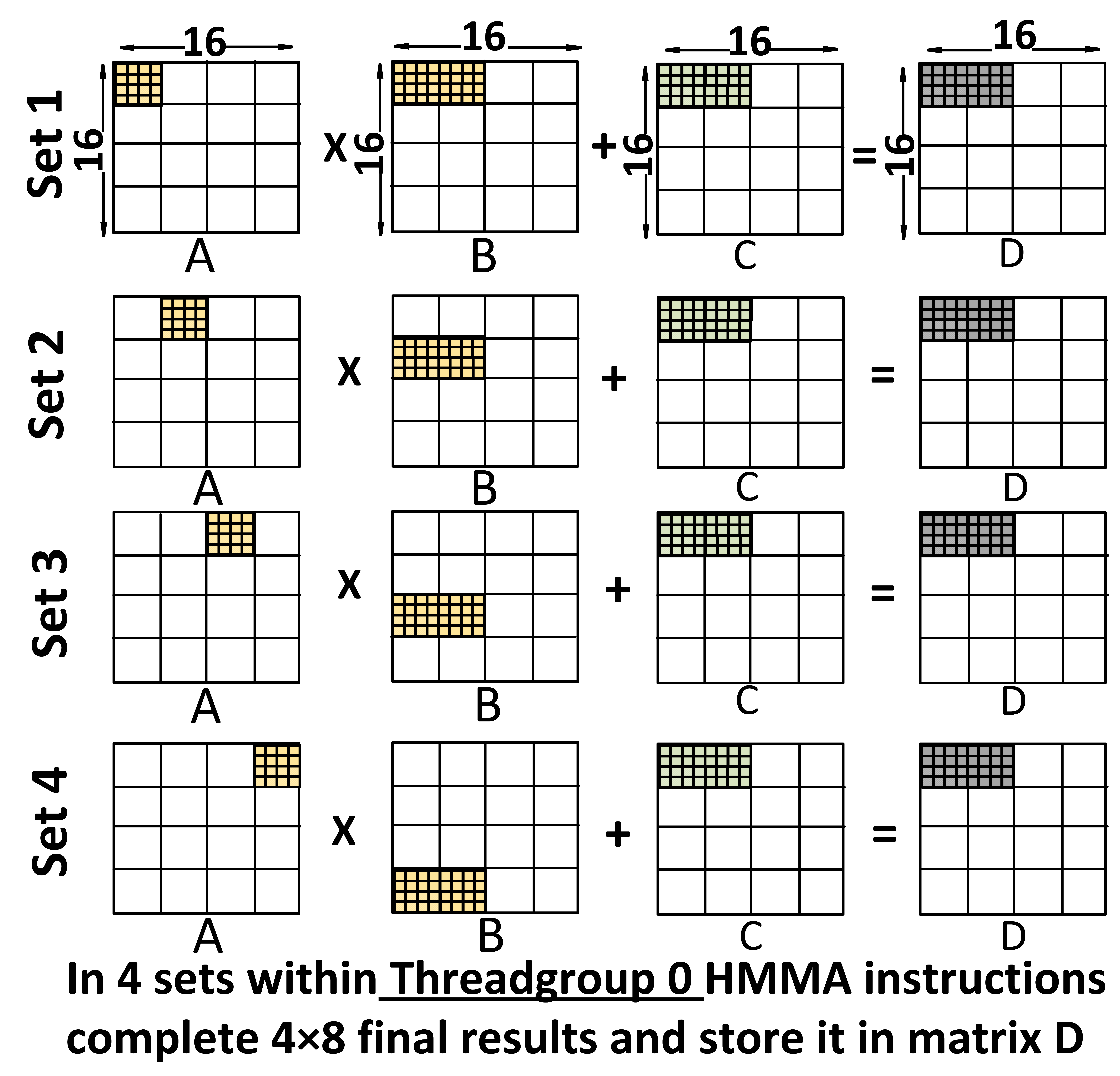}
\caption{Elements accessed in each ``Set''}
\label{set}
\end{subfigure}\vspace{0.1in}
\begin{subfigure}{.48\textwidth}
\centering\includegraphics[width=1\textwidth]{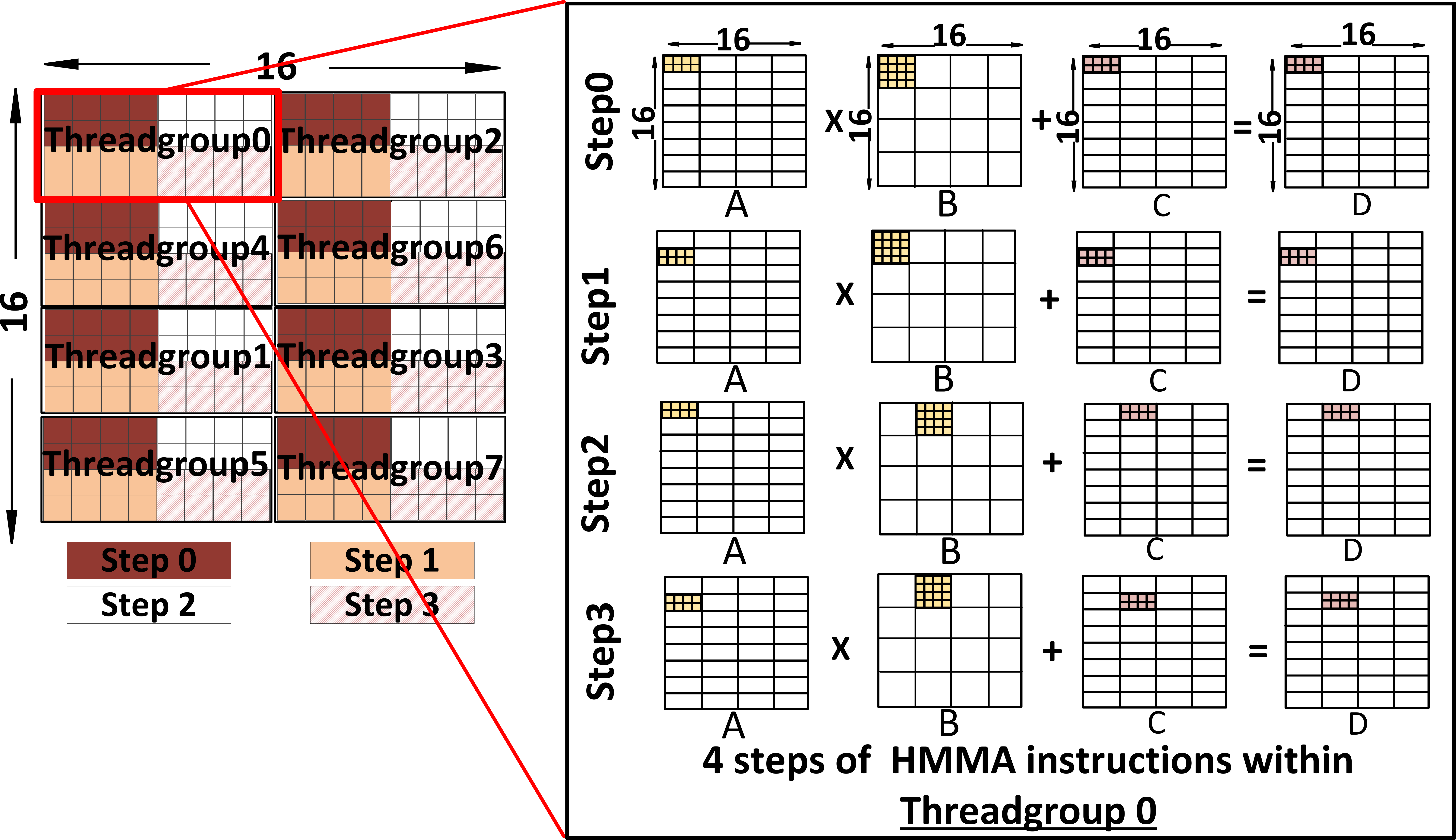}
\caption{Elements accessed in each ``Step'' (mixed-precision mode).}
\label{step}
\end{subfigure}\vspace{0.1in}
\begin{subfigure}{.48\textwidth}
\centering\includegraphics[width=1\textwidth]{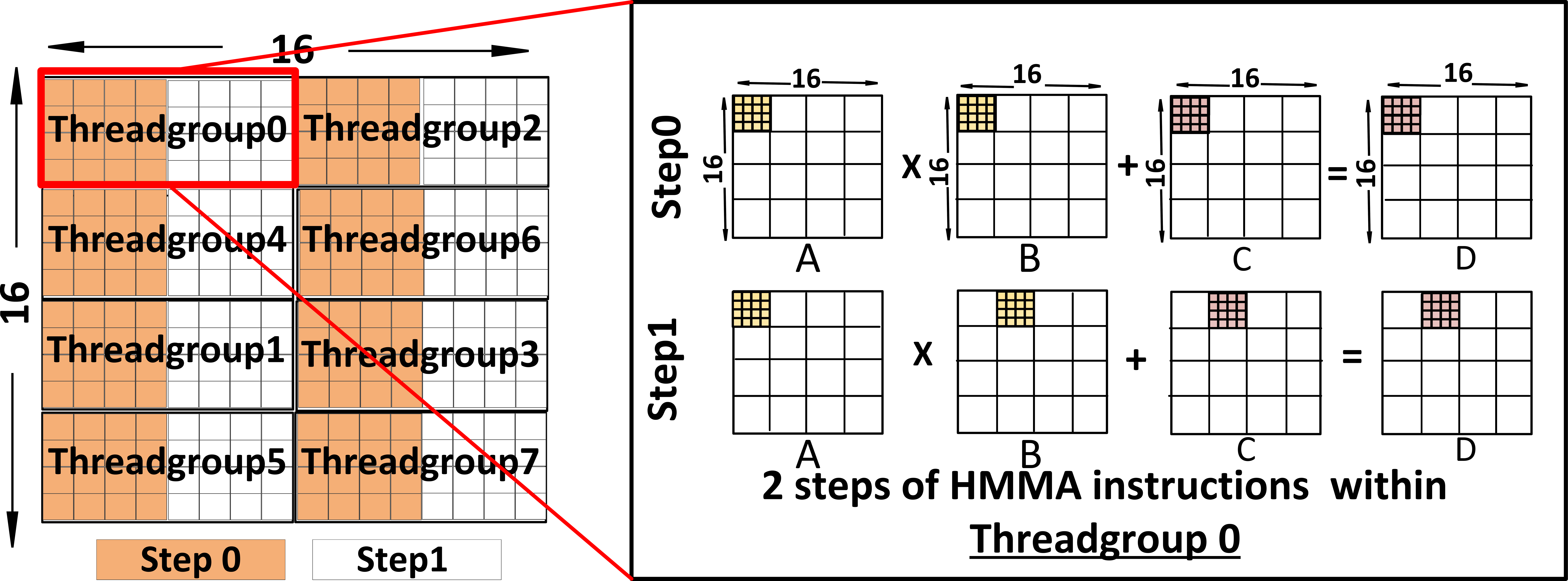}
\caption{Elements accessed in each ``Step'' (FP16 mode).}
\label{step2}
\end{subfigure}\vspace{0.1in}
\caption{\lstinline{HMMA} instruction analysis for Volta (Titan V).}
\end{figure}

We examine the operation of each ``set'' of \lstinline{HMMA} instructions
in Figure~\ref{DisassembledSass}.
As shown in Figure~\ref{set}, 
irrespective of mode, when executing the \lstinline{HMMA} instructions in a set,
each \textit{threadgroup} multiplies a $4\times4$ sub-tile of operand matrix A 
with a $4\times8$ sub-tile of operand matrix B and
accumulates the result with operand matrix C.
For example, when {\em threadgroup}~0 
executes the first set of \lstinline{HMMA} instructions (Set~1)
it multiplies the sub-tile consisting of the first four rows and columns of operand matrix A 
with the sub-tile consisting of the first four rows and first eight columns of operand matrix B.
The result is accumulated with a $4\times8$ sub-tile of operand matrix C and stored in a
$4\times8$ sub-tile of operand matrix~D.

Figure~\ref{step} shows the detailed operation of each \lstinline{HMMA}
``step'' within a ``set'' for threadgroup~0 for mixed-precision mode.  Each
``set'' of \lstinline{HMMA} instructions contains four ``steps''.  We find in
each step, a $2\times4$ sub-tile of operand matrix A is multiplied with a
$4\times4$ sub-tile of operand matrix B, accumulated with a $2\times4$ sub-tile
of operand matrix C. 

Similarly, Figure~\ref{step2} shows the detailed operation of each
\lstinline{HMMA} ``step'' within a ``set'' for threadgroup~0 for FP16 mode.
Each set of \lstinline{HMMA} instructions contains two ``steps''.  In each
step, every threadgroup multiplies a $4\times4$ sub-tile of operand matrix A
with a $4\times4$ sub-tile of operand matrix B and accumulates the result with
matrix C. 

\begin{figure*}
\centering
\begin{subfigure}{.2\textwidth}
\centering\includegraphics[width=1\textwidth]{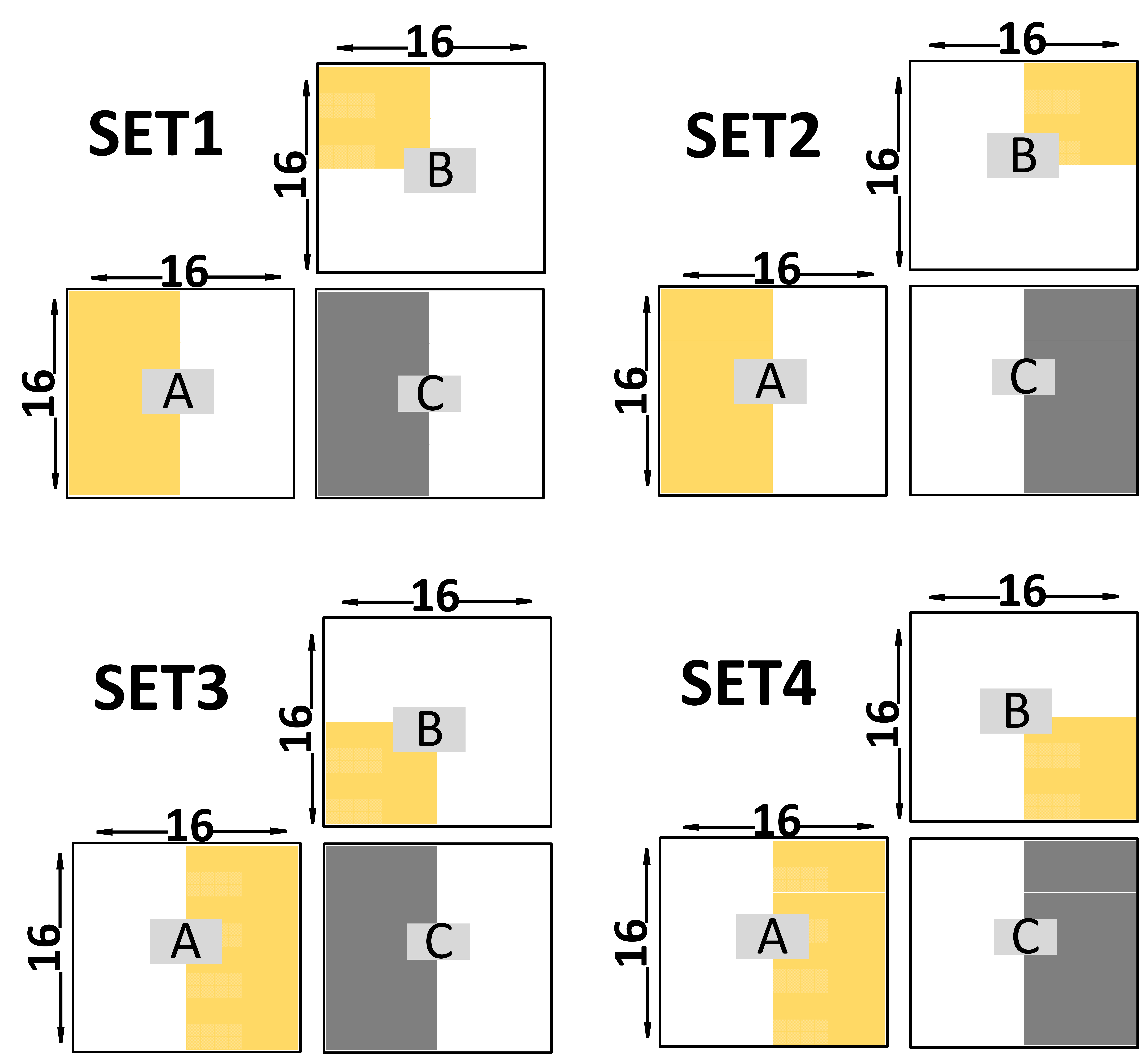}
\caption{Mixed and FP16,\\$16\times16\times16$}
\label{TuringGPU1}
\end{subfigure}\hspace{0.2in}\vspace{0.1in}
\begin{subfigure}{.2\textwidth}
\centering
\centering\includegraphics[width=1\textwidth]{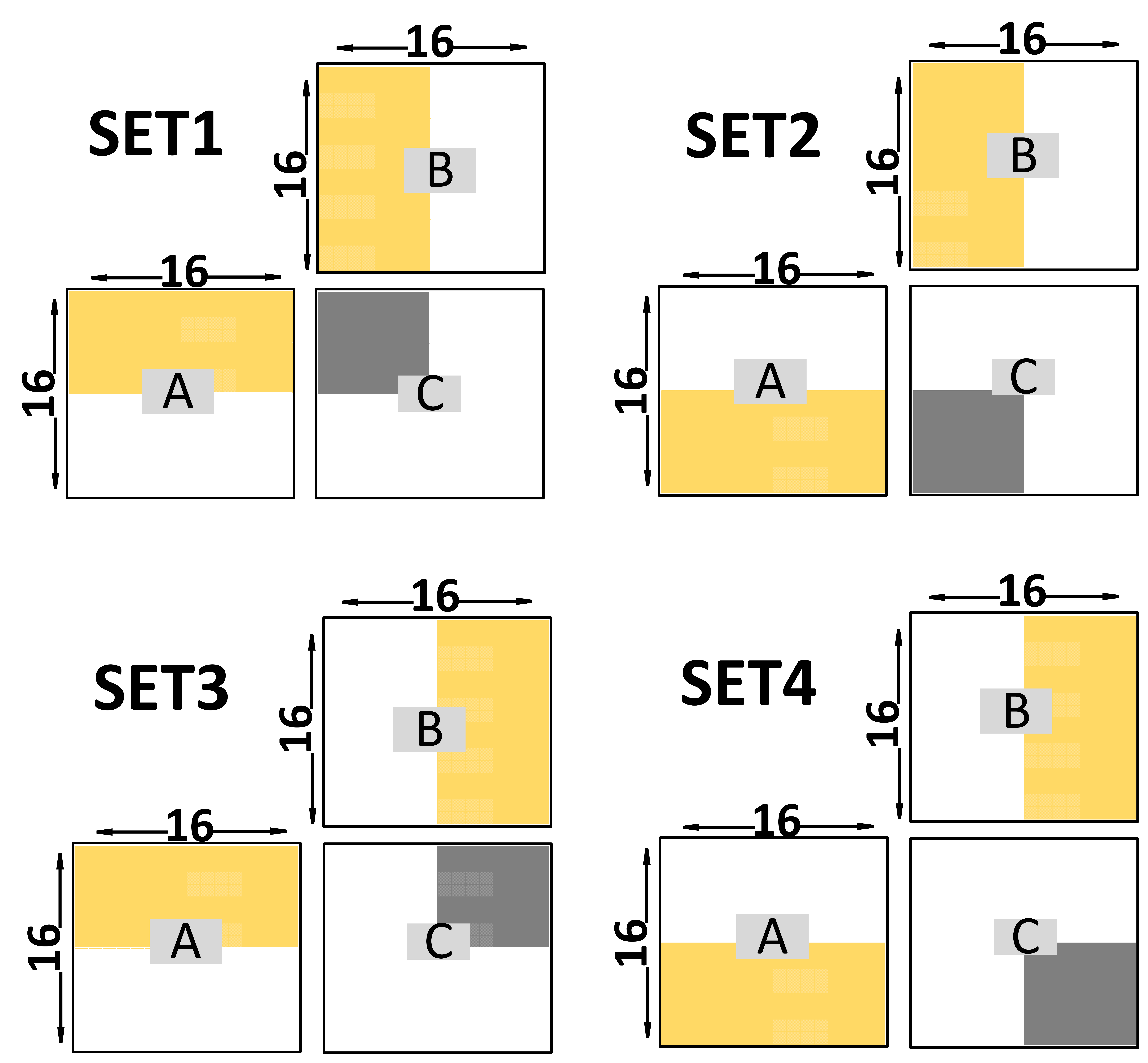}
\vspace{0.02in}
\caption{8-bit, $16\times16\times16$}
\label{TuringGPU2}
\end{subfigure}\hspace{0.2in}\vspace{0.1in}
\begin{subfigure}{.4\textwidth}
\includegraphics[width=1\textwidth]{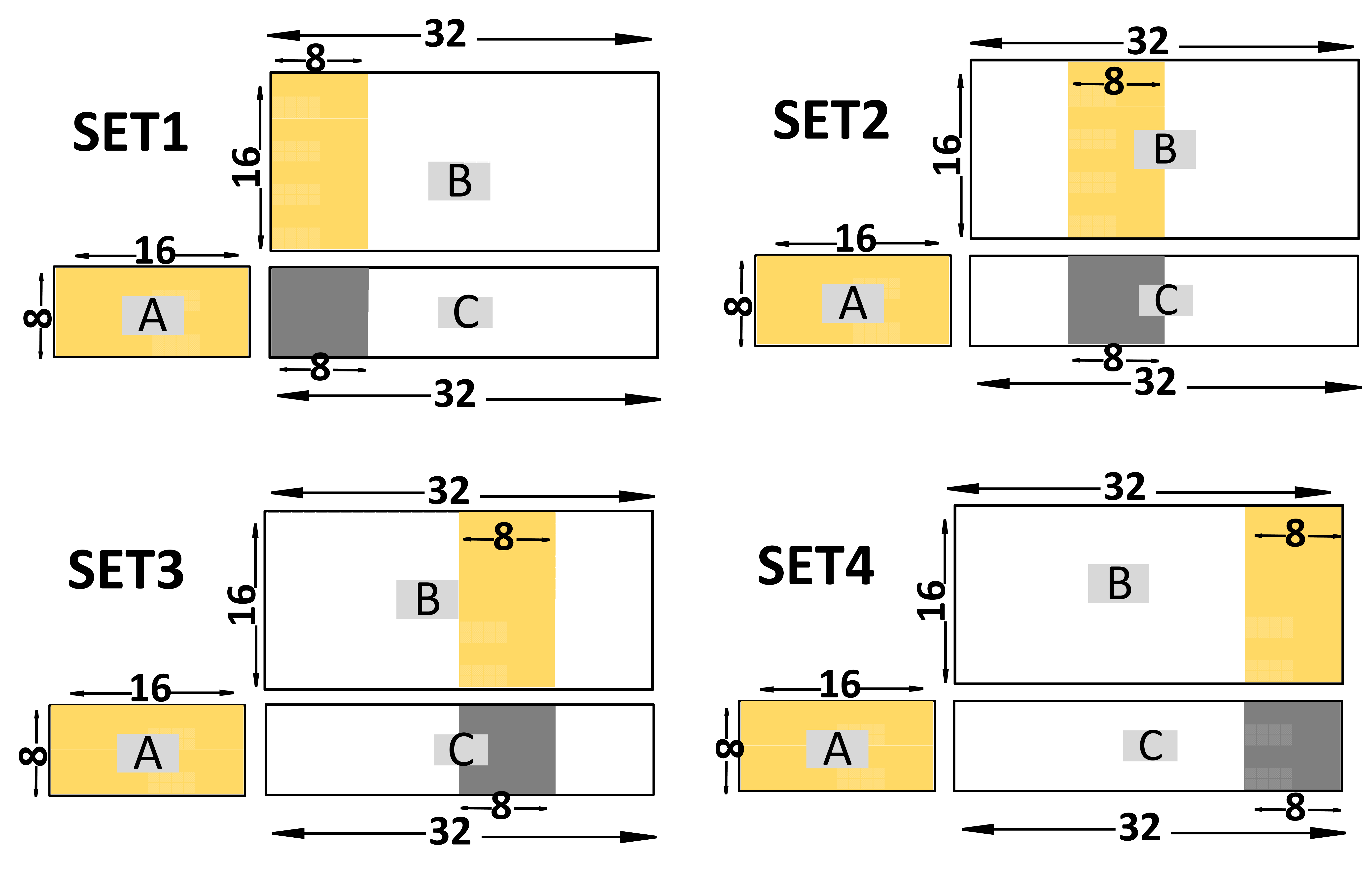}
\caption{8-bit, $8\times32\times16$}
\label{TuringGPU6}
\end{subfigure}\vspace{0.1in}
\begin{subfigure}{.2\textwidth}
\centering
\centering\includegraphics[width=1\textwidth]{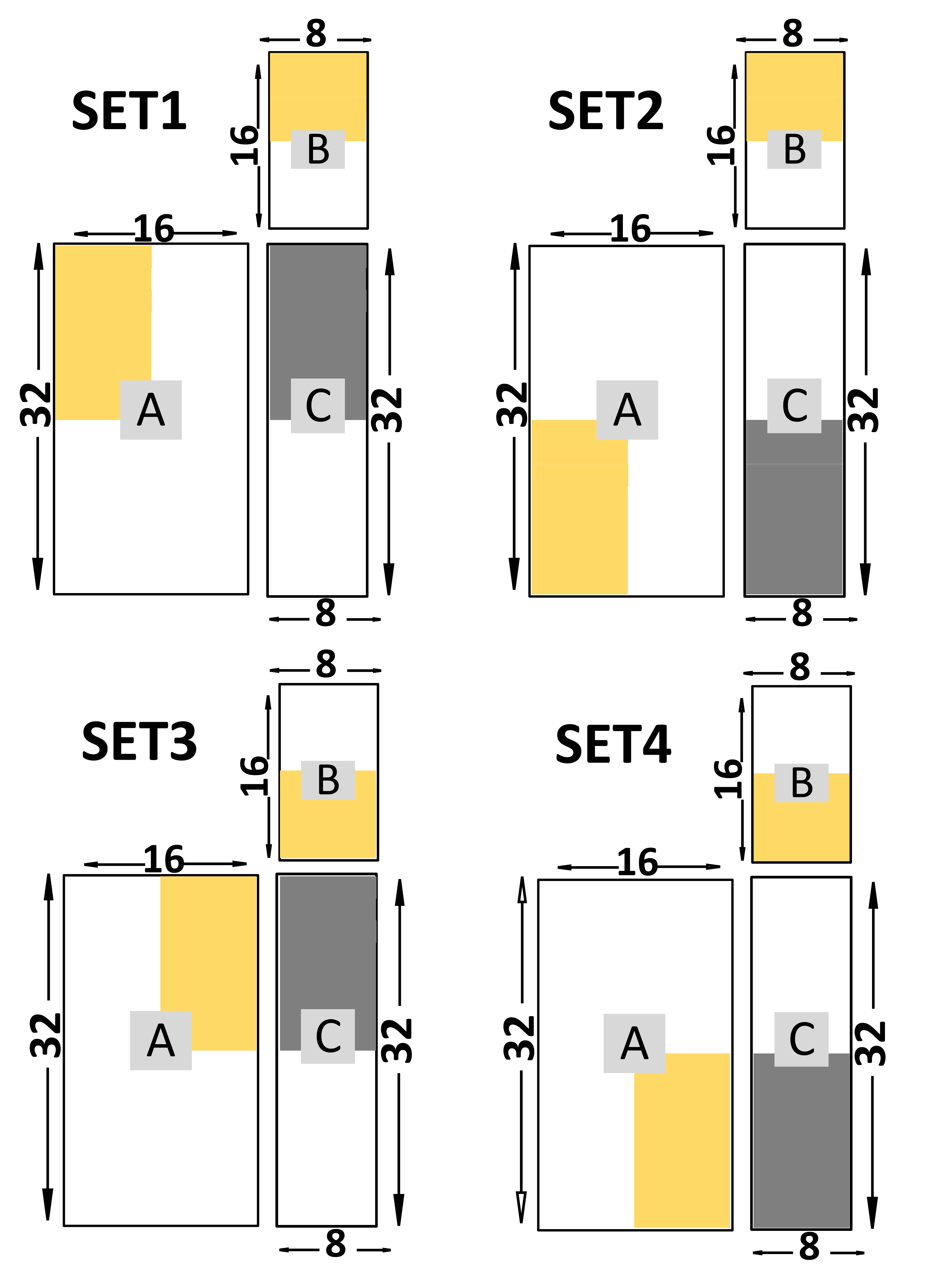}
\caption{Mixed and FP16,\\$32\times8\times16$}
\label{TuringGPU3}
\end{subfigure}\hspace{0.2in}\vspace{0.1in}
\begin{subfigure}{.2\textwidth}
\centering
\centering\includegraphics[width=1\textwidth]{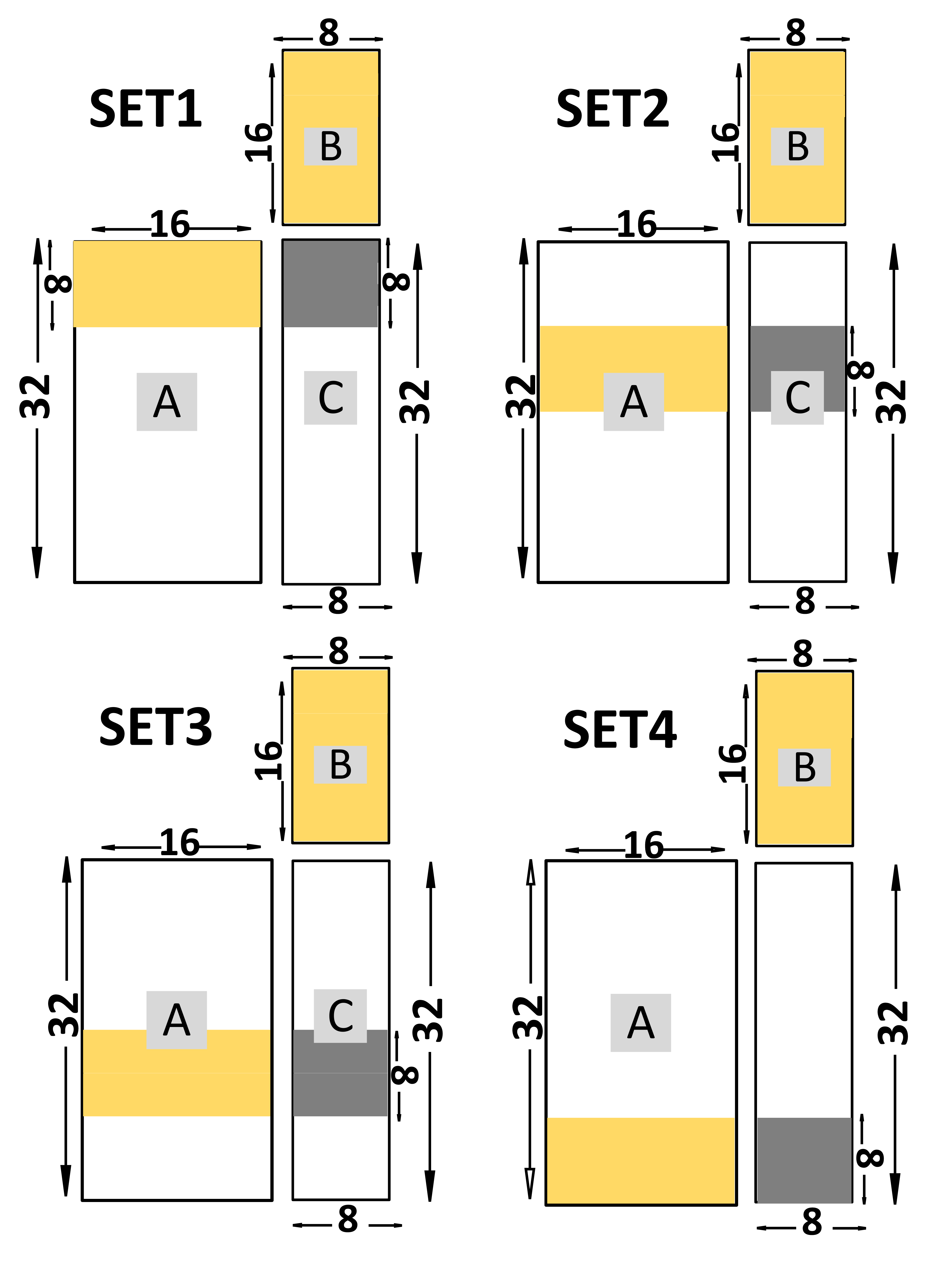}
\vspace{0.02in}
\caption{8-bit, $32\times8\times16$}
\label{TuringGPU4}
\end{subfigure}\vspace{0.1in}\hspace{0.1in}
\begin{subfigure}{0.4\textwidth}
\includegraphics[width=1\textwidth]{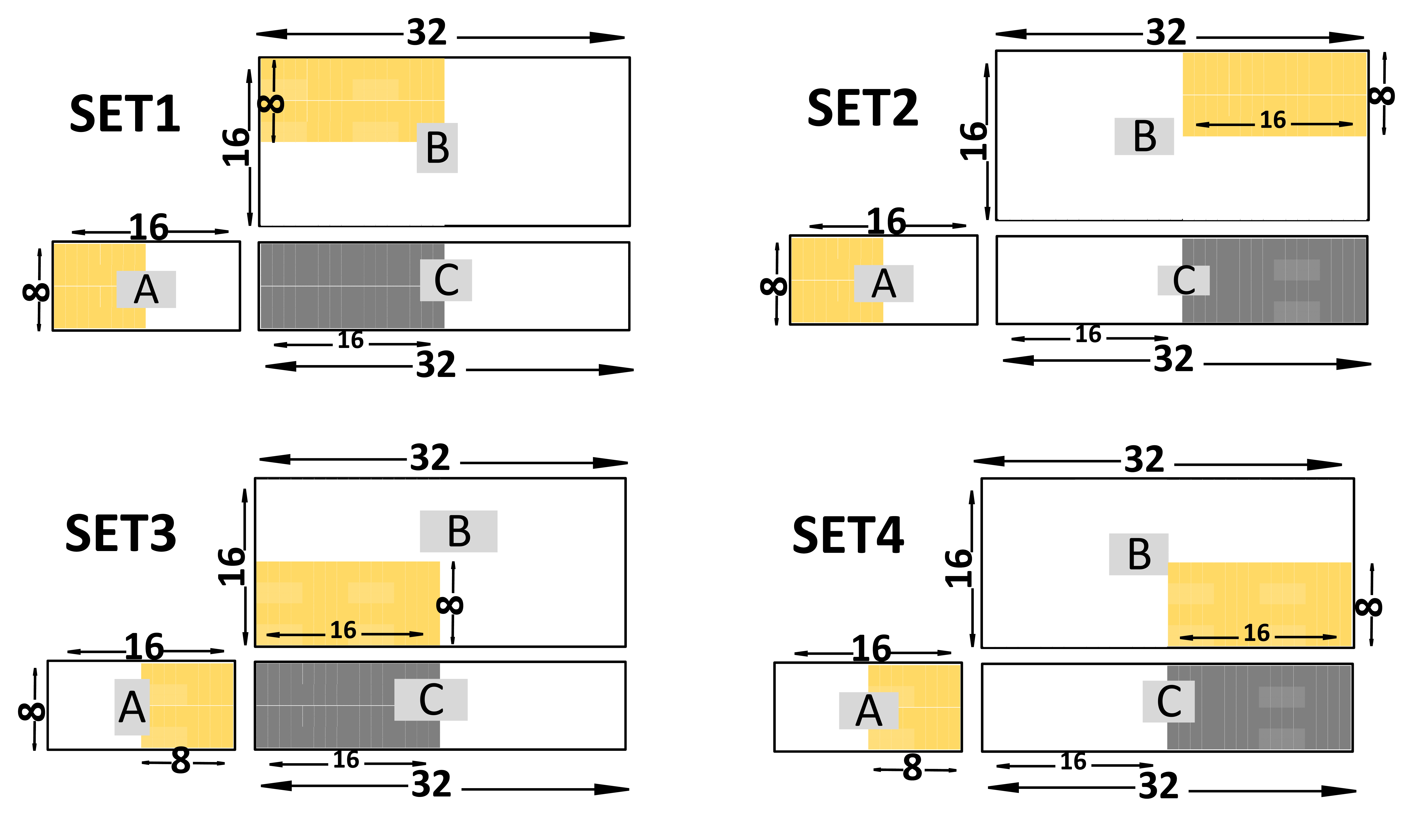}
\caption{Mixed and FP16, $8\times32\times16$.}
\label{TuringGPU5}
\end{subfigure}\hspace{0.1in}\vspace{0.1in}
\caption{\lstinline{HMMA} instruction analysis for Turing (RTX2080).}
\label{TuringGPU}
\end{figure*}

\subsubsection{Turing}

Figure~\ref{TuringGPU} illustrates the elements accessed by \lstinline{HMMA} 
instructions on the Turing GPU architecture.
The  ``step'' annotation found on \lstinline{HMMA} SASS instructions in Volta is not present
in Turing.  Given the latency results in Table~\ref{table:3} do not
suggest increased parallelism one possibility is similar ``steps'' 
are sequenced by the microarchitecture using a state-machine.  
We make the following observations:
\begin{itemize}
 \item The elements accessed for a particular mode are similar for different tile configurations.
 \item In FP16 and mixed precision mode the computation pattern is the product between two subtiles where one of the subtile is $8\times 8$ and the other subtile is either $16\times 8$ or $8\times 16$. For example, for tile size $16\times 16\times 16$ or $32\times 8\times 16$, the computation in SET~1 is the product between the $16\times 8$ subtile of matrix A with the $8\times 8$ subtile of matrix B whereas for tile size $8\times 32\times 16$  the product is between the $8\times 8$ subtile of matrix A with the $8\times 16$ subtile of matrix B. 
 \item For the 8-bit mode the computation pattern is the product between the $8\times16$ subtile of matrix A with $16\times8$ subtile of matrix B. 
  \item In 4-bit mode, each \lstinline{wmma.mma} PTX instruction is implemented with a single \lstinline{HMMA} SASS instruction so we omit 4-bit mode in Figure~\ref{TuringGPU}.
\end{itemize}

\subsection{Discussion}

\begin{figure*}
\centering
\begin{subfigure}{.3\textwidth}
\centering\includegraphics[width=0.8\textwidth]{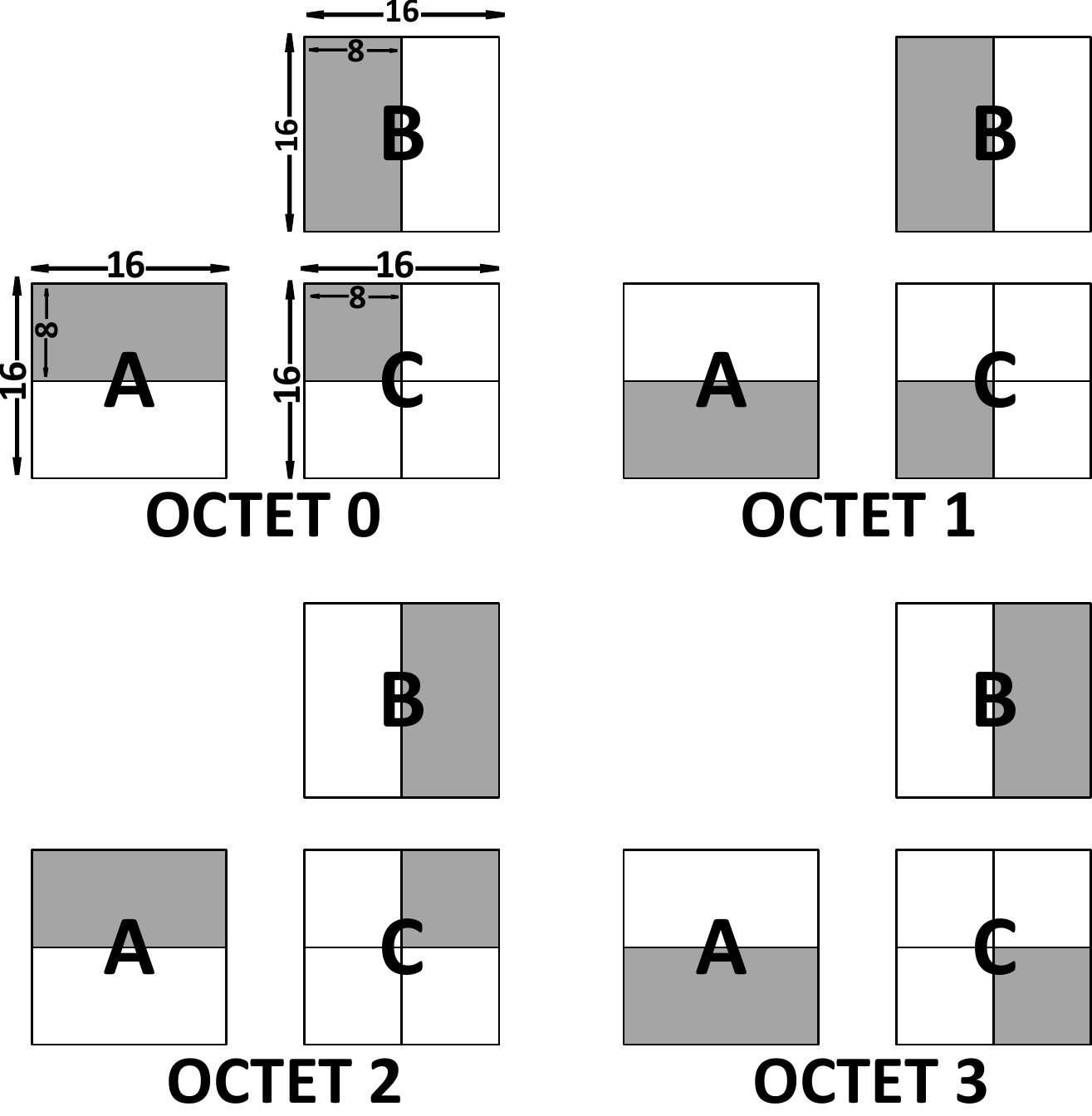}
\caption{Elements of operand matrices accessed by each \workgroup}
\label{Workgroup}
\end{subfigure}\hspace{0.1in}
\begin{subfigure}{.3\textwidth}
\centering
\centering\includegraphics[width=.75\textwidth]{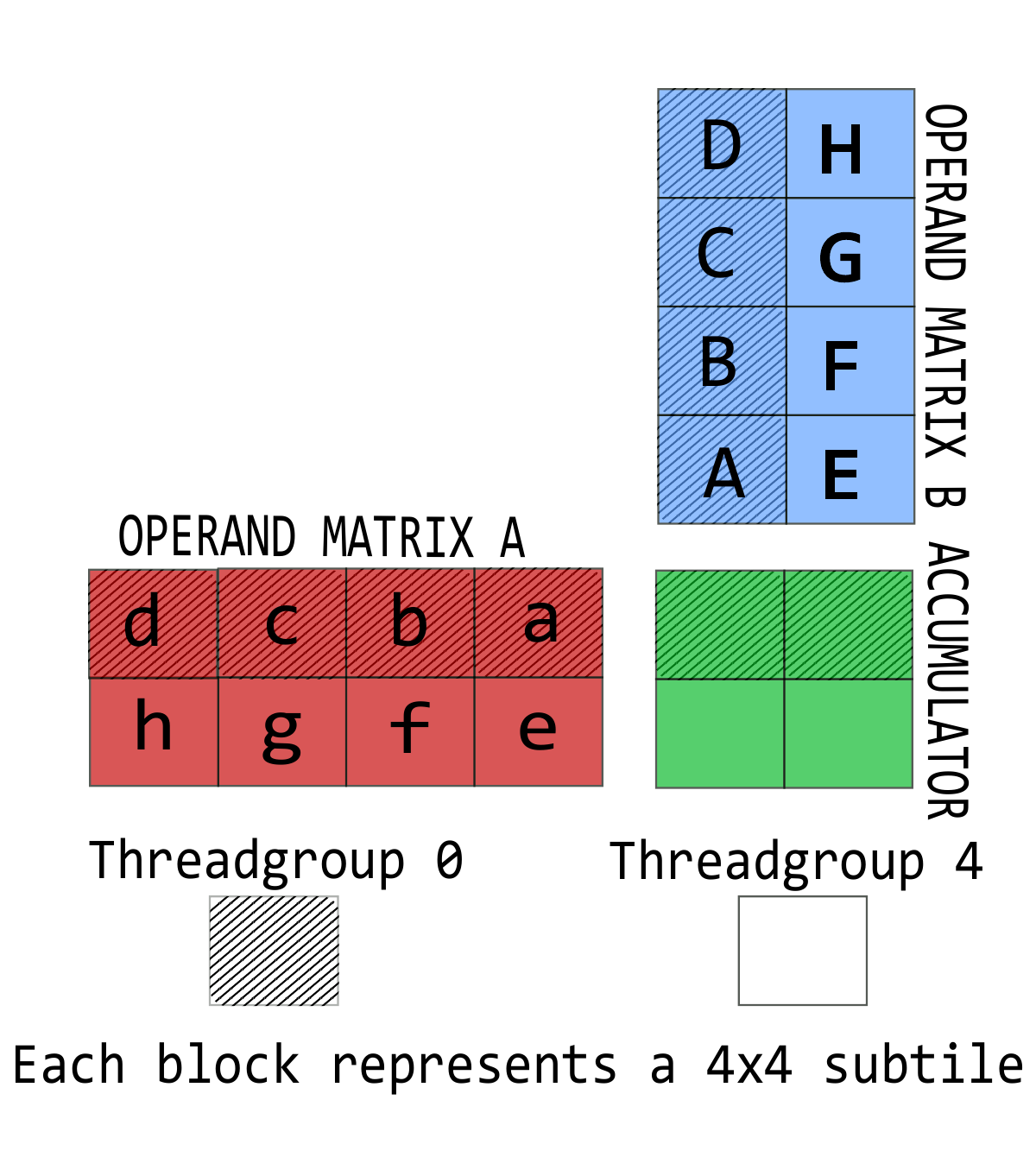}
\caption{Outer product formulation during sets and steps in an \workgroup}
\label{explanation}
\end{subfigure}\hspace{0.1in}
\begin{subfigure}{.3\textwidth}
\centering
\centering\includegraphics[width=1\textwidth]{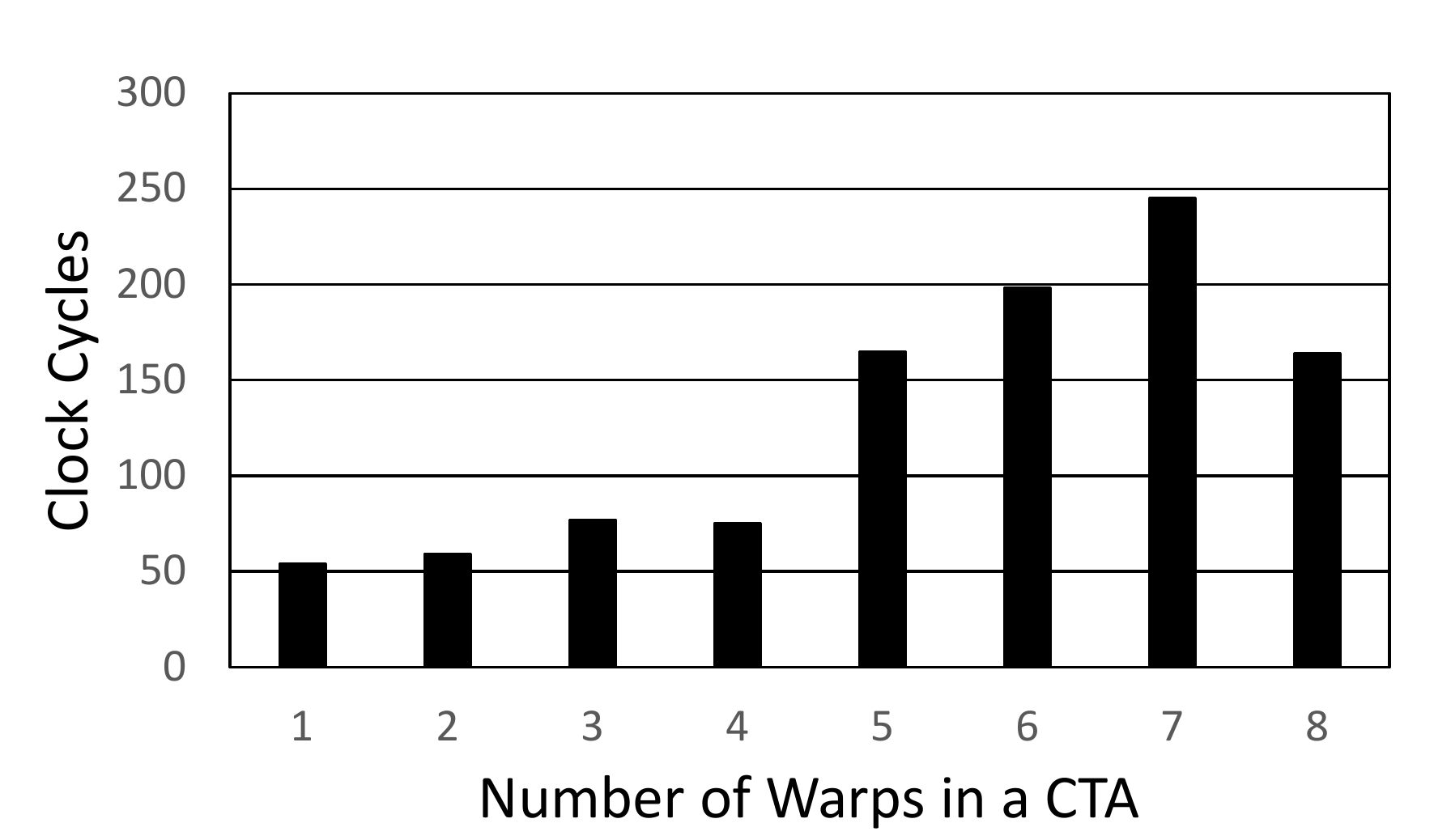}
\caption{Cycles to execute parallel HMMA operations versus number of warps per SM}
\label{CTA}
\end{subfigure}\hspace{0.1in}
\caption{}
\label{LL}
\end{figure*} 

In this section, we provide our analysis of the results presented above for Volta and
infer a possible rationale for why execution is broken into ``sets'' and ``steps''.

Recall, each element of the input matrix is loaded by two different \textit{threadgroups}.
We wrote a microbenchmark to help determine how the fragments loaded by different threads are used by a \lstinline{HMMA} instruction.
For example, to determine how operand matrix elements loaded by thread~0 are used, we altered
these values and observed how the result is affected. 
We found that \textit{threadgroups} work in pairs to compute
$8\times8$ subtiles of the result.  We call each such pair of threadgroups an \workgroup.
There are four \workgroups\ in a warp.

Table~\ref{table:1} shows the pair of \textit{threadgroup} constituting each
\textit{octet}, which in general can be formulated as \workgroup\ X
= {\textit{threadgroup} X $\bigcup$ \textit{threadgroup} X+4 } where X lies in
between 0 to 3. Table \ref{table:1} also uses the notation [Row\_Start :
Row\_End, Col\_Start : Col\_End ] to show the subtile of the operand matrix A
and B accessed by the threads inside each \workgroup. The elements loaded
by the \workgroup\ remain the same irrespective of the layout in which
the operand matrices are stored. 

Table~\ref{table:1} shows each element of the operand matrices A and B is
loaded twice by threads in a different \textit{threadgroup}.
This enables each
\workgroup\ to work independently.  Specifically, each \workgroup\ reads
an $8\times 16$ subtile of operand matrix A,
an $16\times 8$ subtile of operand matrix B
and an $8\times 8$ subtile of operand matrix C as shown in Figure~\ref{Workgroup}.

To better understand the organization of threads into \workgroups, we 
analyzed the calculation performed by \workgroups\ in different ``sets'' and ``steps''.
As shown in Figure~\ref{explanation},
in each set, every \workgroup\ performs the outer product between input subtiles.
For example, in Set~1 the outer
product between input subtile $[a],[e]$ and $[A],[E]$ is completed to generate
the partial result $[aA], [aE], [eA]$ and $[eE]$. Here each $[a],[e],[A],[E]$
represents a $4 \times 4$ subtile.  To compute 
$[aE]$,  \textit{threadgroup}~0 needs operand matrix B subtile $[E]$ which is only
loaded by \textit{threadgroup}~4.  Similarly, to compute $[eA]$,
\textit{threadgroup}~4 needs operand matrix B subtile $[A]$ which is only loaded by
\textit{threadgroup}~0.  Thus, while \textit{threadgroups} cannot,
\workgroups\ {\em can} work independently.  Table~\ref{table:2}
expands upon Figure~\ref{explanation} to tabulate all the outer product computations
performed in different sets and steps.

\begin{table}[h]
 \centering
\begin{tabular}{|p{0.06\textwidth}|p{0.1\textwidth}|p{0.1\textwidth}|p{0.1\textwidth}|}
 \hline
 \rule{0pt}{0.04\textwidth} \Workgroup & Threadgroup  & Matrix A & Matrix B \\
  \hline
  \hline
   0 &  0 and 4  & [0:7,0:15] & [0:15,0:7] \\ 		
  \hline
   1 &  1 and 5  & [8:15,0:15] &[0:15,0:7] \\ 		
  \hline
   2 &  2 and 6  & [0:7,0:15] &[0:15,8:15] \\ 	
  \hline
   3 &  3 and 7  & [8:15,0:15] & [0:15,8:15] \\  	
 \hline
\end{tabular}
\caption{\Workgroup\ composition and elements accessed}\vspace{1pt}
\label{table:1}
\end{table}

\begin{table}[h]
 \centering
\begin{tabular}{ |c|c|c|c|} 
 \hline
  \rule{0pt}{15pt}  SET & STEP & Threadgroup X & Threadgroup X+4 \\
  \hline
  \hline
   \multirow{4}{*}{1}&  0 & $a[0:1]\times A$  &  $e[0:1]\times A$ \\ 		
     				  &  1 & $a[2:3]\times A$  &  $e[2:3]\times A$ \\ 		
      				  &	 2 & $a[0:1]\times E$  &  $e[0:1]\times E$ \\ 	
      				  &	 3 & $a[2:3]\times E$  & $e[2:3]\times E$ \\  	
 \hline
    \multirow{4}{*}{2}&  0 & $b[0:1]\times B$  &  $f[0:1]\times B$ \\ 		
     				  &  1 & $b[2:3]\times B$  &  $f[2:3]\times B$ \\ 		
      				  &	 2 & $b[0:1]\times F$  &  $f[0:1]\times F$ \\ 	
      				  &	 3 & $b[2:3]\times F$  & $f[2:3]\times F$ \\  	
 \hline
    \multirow{4}{*}{3}&  0 & $c[0:1]\times C$  &  $g[0:1]\times C$ \\ 		
     				  &  1 & $c[2:3]\times C$  &  $g[2:3]\times C$ \\ 		
      				  &	 2 & $c[0:1]\times G$  &  $g[0:1]\times G$ \\ 	
      				  &	 3 & $c[2:3]\times G$  & $g[2:3]\times G$ \\  	
 \hline
    \multirow{4}{*}{4}&  0 & $d[0:1]\times D$  &  $h[0:1]\times D$ \\ 		
     				  &  1 & $d[2:3]\times D$  &  $h[2:3]\times D$ \\ 		
      				  &	 2 & $d[0:1]\times H$  &  $h[0:1]\times H$ \\ 	
      				  &	 3 & $d[2:3]\times H$  & $h[2:3]\times H$ \\  	
 \hline
\end{tabular}
\caption{\Workgroup\ computation details}\vspace{0pt}
\label{table:2}
\end{table}\vspace{0pt}
 
\section{A Tensor Core Microarchitecture}

\begin{figure*}
\centering
\begin{subfigure}{.9\textwidth}
\centering\includegraphics[width=1\textwidth]{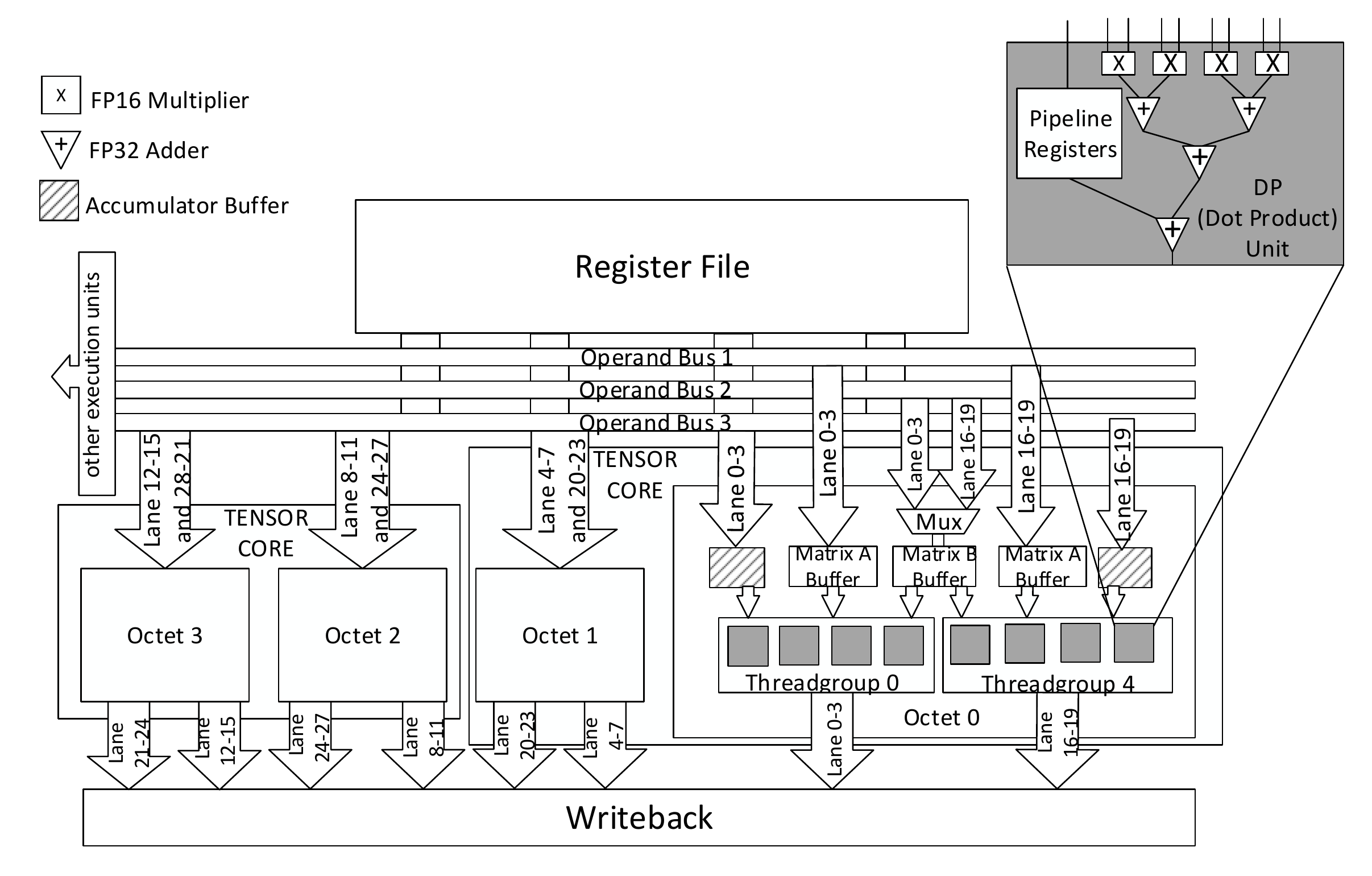}
\end{subfigure}\vspace{10pt}
\caption{Proposed Tensor Core Microarchitecture}
\label{TC Architecture}
\end{figure*}

In this section we present a tensor core microarchitecture consistent with the
observations made for Volta earlier in the paper.

Recall each tensor core completes a $4\times4$ matrix-multiply and accumulate
each cycle.  To achieve this, each tensor core must be able to perform sixteen
four-element dot-products (FEDPs) each cycle.  As shown in
Figure~\ref{DisassembledSass} and \ref{step}, in steady state, a
\textit{threadgroup} takes two cycles to generate a $2\times 4$ subtile of the
output matrix. Thus, across all threads in a warp a HMMA instruction is executing 
32 FEDP per cycle.  Since each tensor core can only complete 16 FEDP per cycle it
follows that full throughput requires two tensor cores per sub-core within an SM.
To confirm this we wrote a microbenchmark that repeatedly executes HMMA operations, varies the
number of warps per thread block and the number of thread blocks executing
concurrently constant.
As shown in Figure~\ref{CTA}, this microbenchmark shows that only four warps can
concurrently execute on a single SM, but the Titan~V SM has 8 tensor cores per SM. 
Thus, each warp appears to utilize two tensor cores.

Next, we consider register access bandwidth.  The data in
Figure~\ref{DisassembledSASS_32} suggests the minimum initiation interval of an
HMMA instruction is two cycles.  There are three source operands and as noted 
in Section~\ref{sec:sass} for each source operand a pair of 32-bit registers is read.
Taking all these factors into account the
total register fetch bandwidth is $32\times2\times3\times32 =$ 6kb every 2~cycles per
warp.  This bandwidth is sufficient for a warp to fetch the following every two cycles:
eight $2\times4$ FP16 subtiles for operand A, 
eight $4\times4$ FP16 subtiles for operand B, and 
eight $2\times4$ FP32 subtiles or eight $4\times4$ FP16 subtiles for operand C.
Given every warp accesses two tensor cores, the register bandwidth per tensor core is 
1.5kb per warp per clock cycle.

NVIDIA states that in Volta INT and FP32 instructions can be co-issued~\cite{Volta}.  
On the other hand tensor core operations reportedly cannot be
co-issued with integer and floating-point arithmetic instructions~\cite{insidevolta}.
We believe the reason is that the tensor cores may be using the register file access ports 
associated with the INT and FP32 cores. 
There are 64 INT and 64 FP32 ALUs inside Titan V SM for a
total of 128 ALUs.  With eight tensor cores inside an SM 
sharing access to the register file each tensor core should be able to access
$\frac{128}{8}\times32=16\times32=512$ bits per source operand per cycle.
Assuming three source operands per ALU (to support multiply-accumulate operations)
this means each tensor core can access 1.5kb/cycle.

Figure~\ref{TC Architecture} illustrates our proposed tensor core microarchitecture. 
Each warp utilizes two tensor cores.  
We assume two \workgroups\ within a warp access each tensor core. 
Sixteen SIMD lanes are dedicated to each tensor core, eight to each \workgroup, and four to
each \textit{threadgroup}. 
Each \textit{threadgroup} lane fetches the operands into internal buffers.
For operand matrix~A and C, each \textit{threadgroup} fetches the operands to its separate buffer whereas
for operand matrix~B both the \textit{threadgroups} fetches to a shared buffer. 
The mode of operation and steps determine the \textit{threadgroup} lane from which
each operand is fetched. The buffers feed sixteen FP16 FEDP units. Inside each FEDP unit, 
multiplication is performed in parallel in the first stage and accumulation occurs over three stages 
for a total of four pipeline stages. As each tensor core consists of sixteen FP16 FEDP units, it is 
capable of completing one $4\times4$ matrix multiplication each cycle.

\section{Modeling and Evaluation}
\subsection{Modelling Tensor Cores} 
\begin{figure*}
\centering
\begin{subfigure}{.3\textwidth}
\centering\includegraphics[width=1\textwidth]{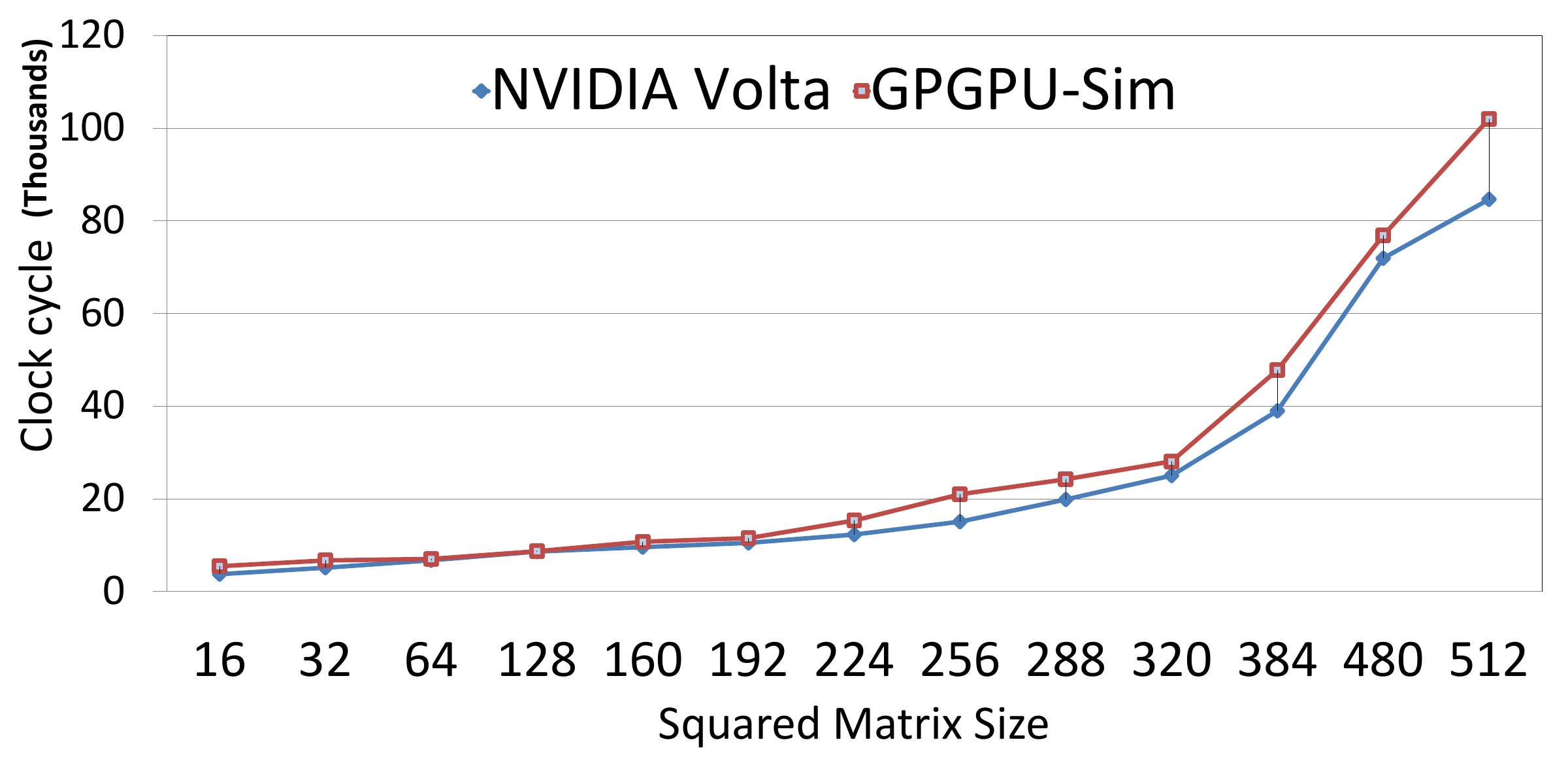}
\caption{WMMA-based GEMM kernel cycle count as matrix size varies.}
\label{modelling}
\end{subfigure}\hspace{0.1in}
\begin{subfigure}{.3\textwidth}
\centering\includegraphics[width=1.0\textwidth]{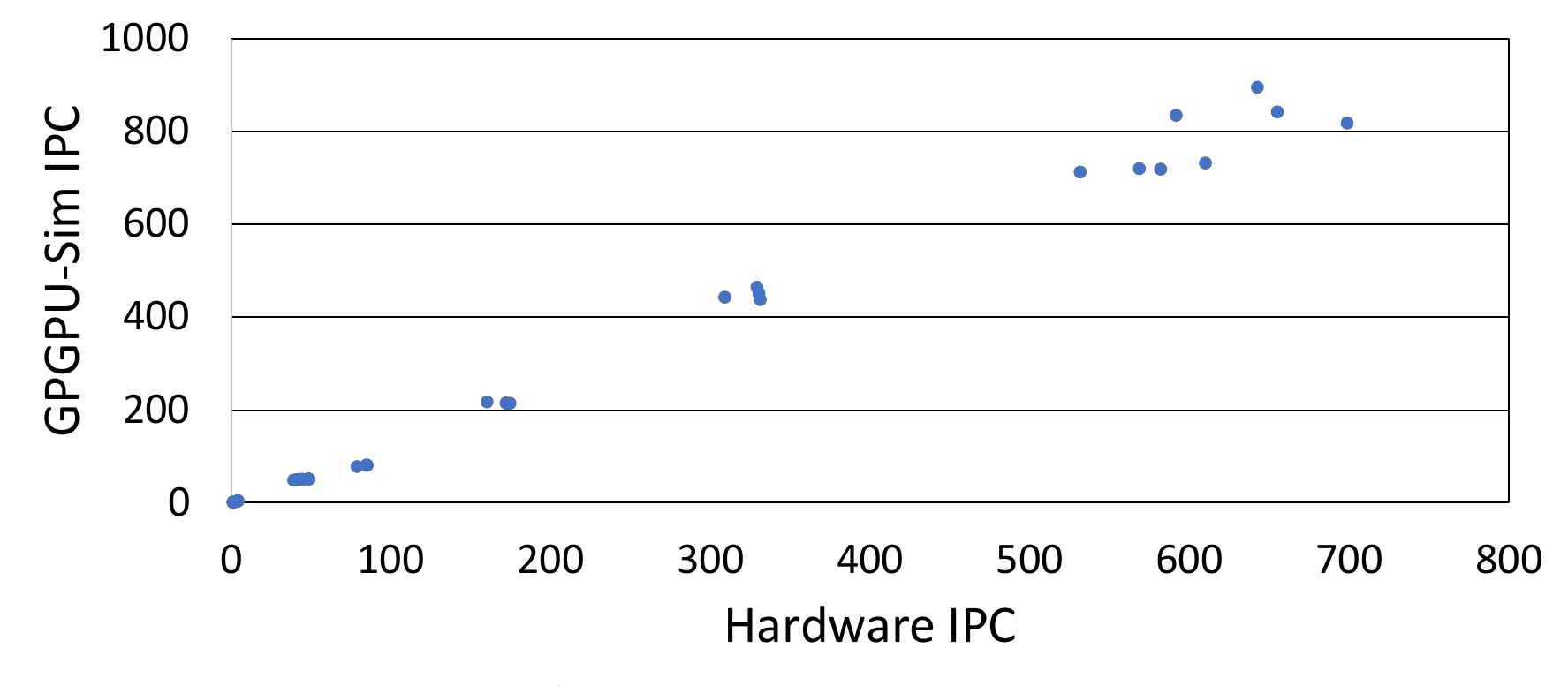}
\caption{Instructions per cycle (IPC) correlation of CUTLASS GEMM kernel on GPGPU-Sim vs Titan~V.} 
\label{cutlass:a}
\end{subfigure}\hspace{0.1in}
\begin{subfigure}{.3\textwidth}
\centering\includegraphics[width=1.0\textwidth]{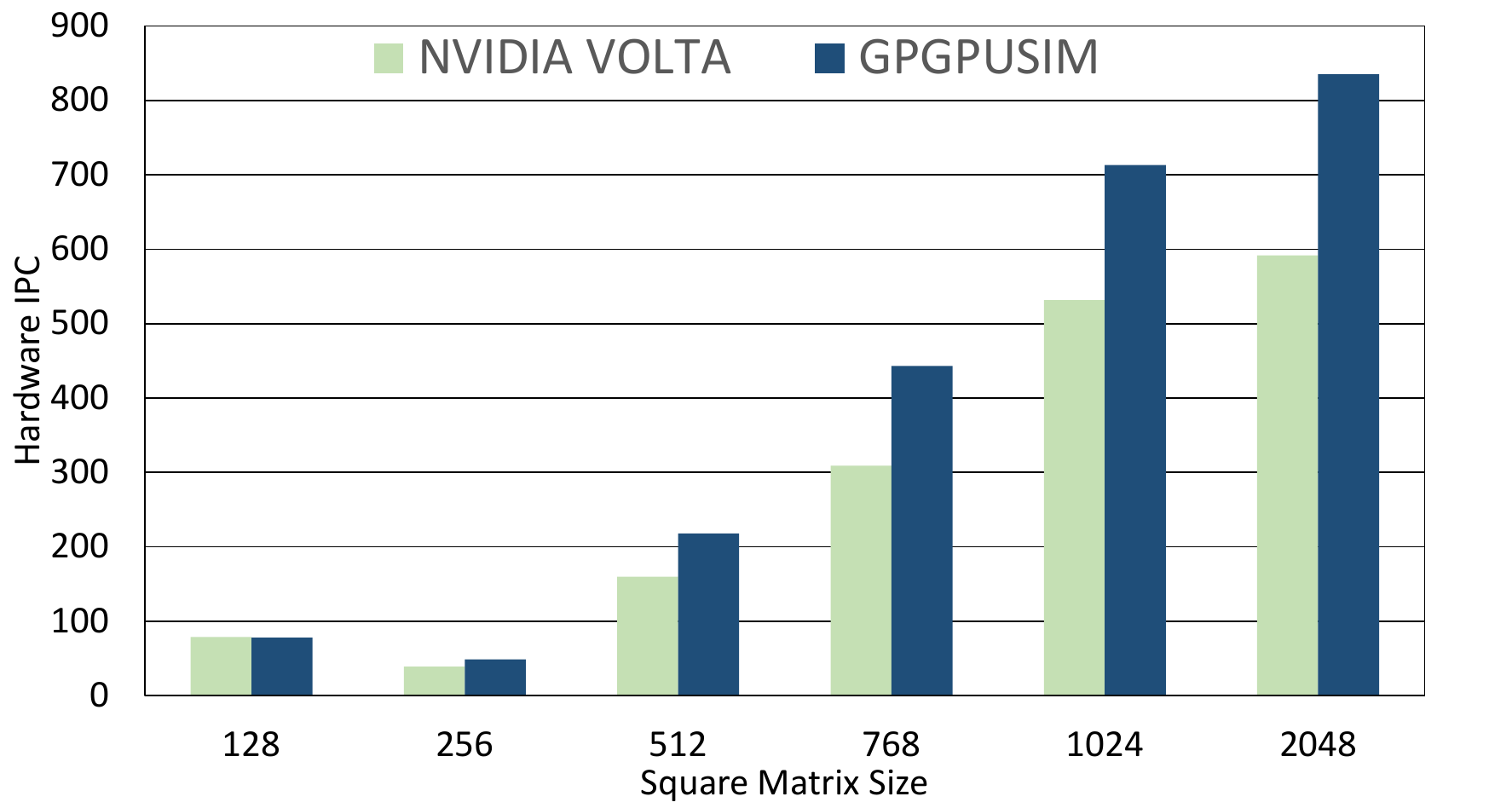}
\caption{CUTLASS-based GEMM kernel cycle count as matrix size varies.}
\label{cutlass:b}
\end{subfigure}\hspace{0.1in}
\caption{Comparison of simulated and actual performance}
\end{figure*}
\begin{figure*}
\centering
\begin{subfigure}{.3\textwidth}
\centering\includegraphics[width=1\textwidth]{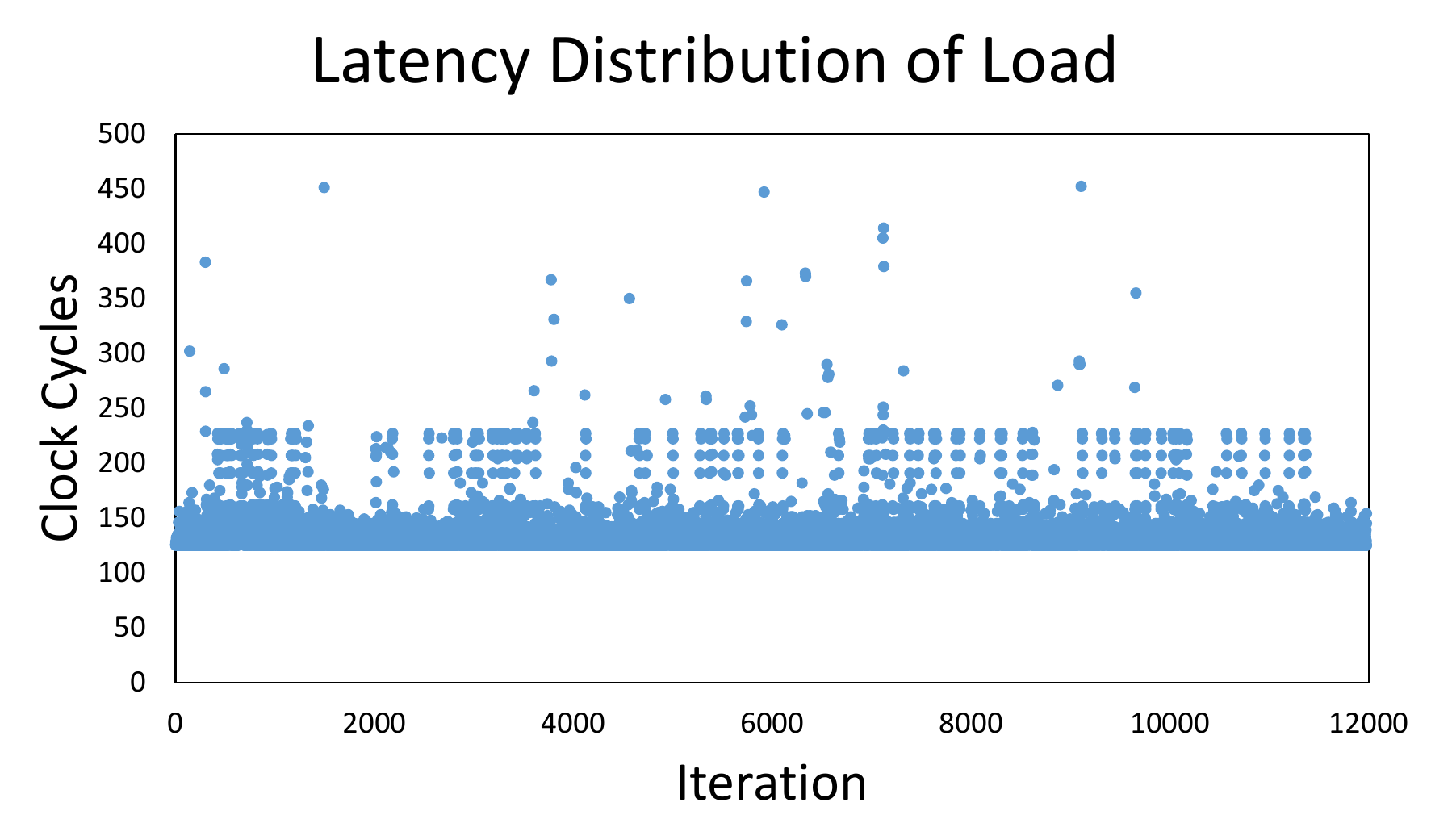}
\end{subfigure}\hspace{0.1in}
\begin{subfigure}{.3\textwidth}
\centering\includegraphics[width=1\textwidth]{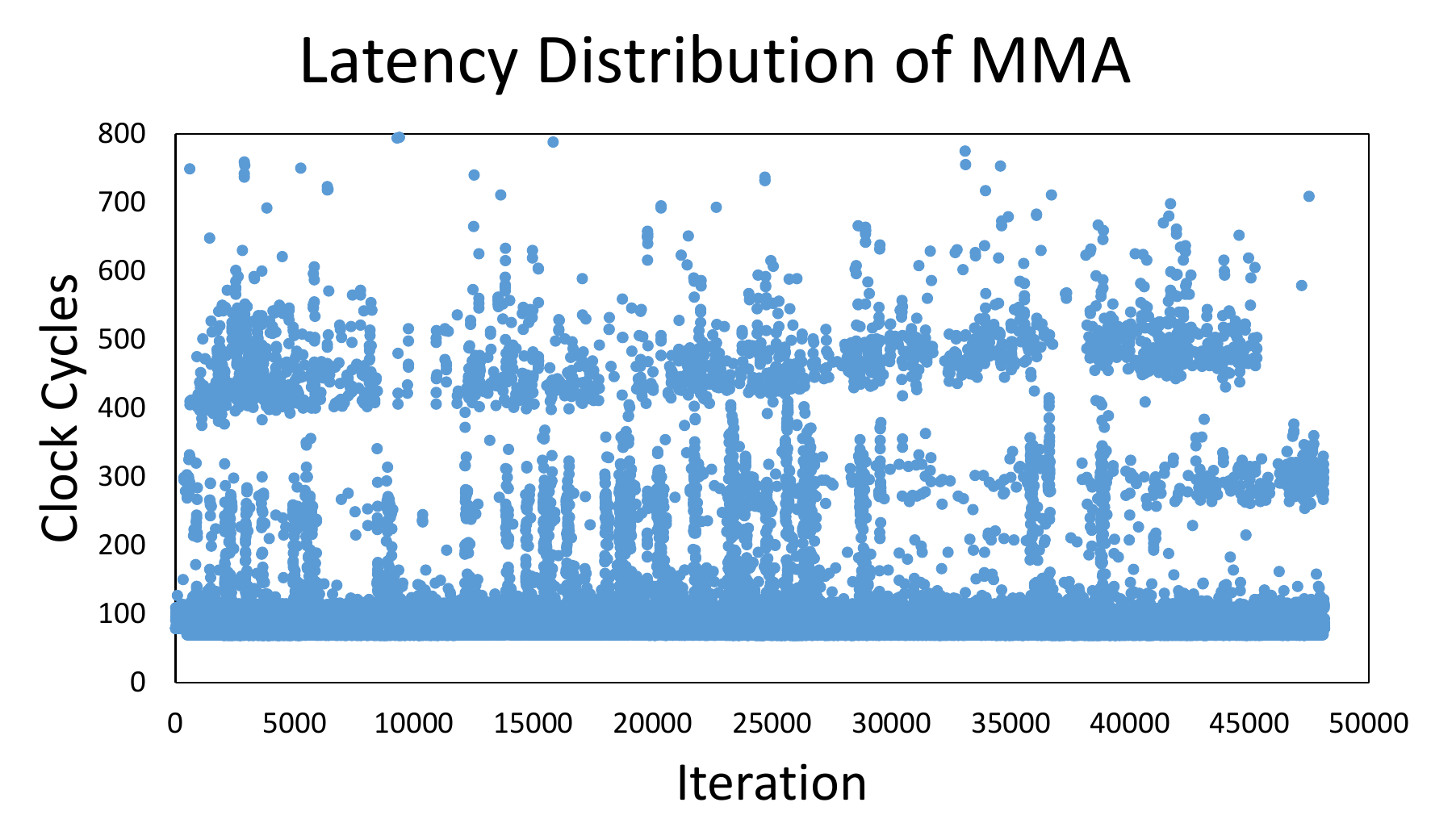}
\end{subfigure}\hspace{0.1in}
\begin{subfigure}{.3\textwidth}
\centering\includegraphics[width=1\textwidth]{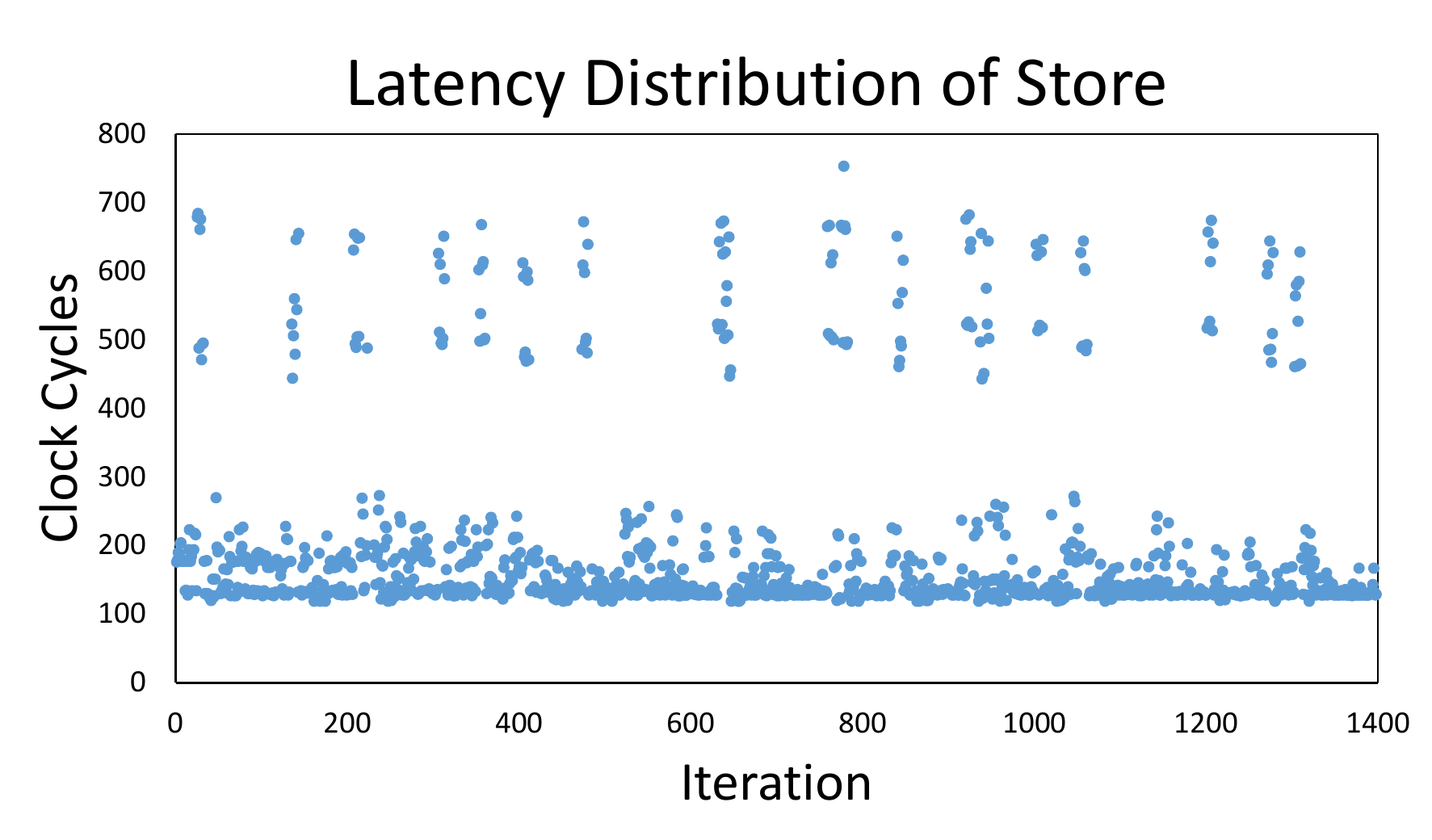}
\end{subfigure}\hspace{0.1in}
\caption{Distribution of \lstinline{wmma.load}, \lstinline{wmma.mma} and \lstinline{wmma.store} latency for matrix size $1024\times1024$ GEMM using shared memory}
\label{latencydist}
\vspace{0pt}
\centering
\begin{subfigure}{.3\textwidth}
\centering\includegraphics[width=1\textwidth]{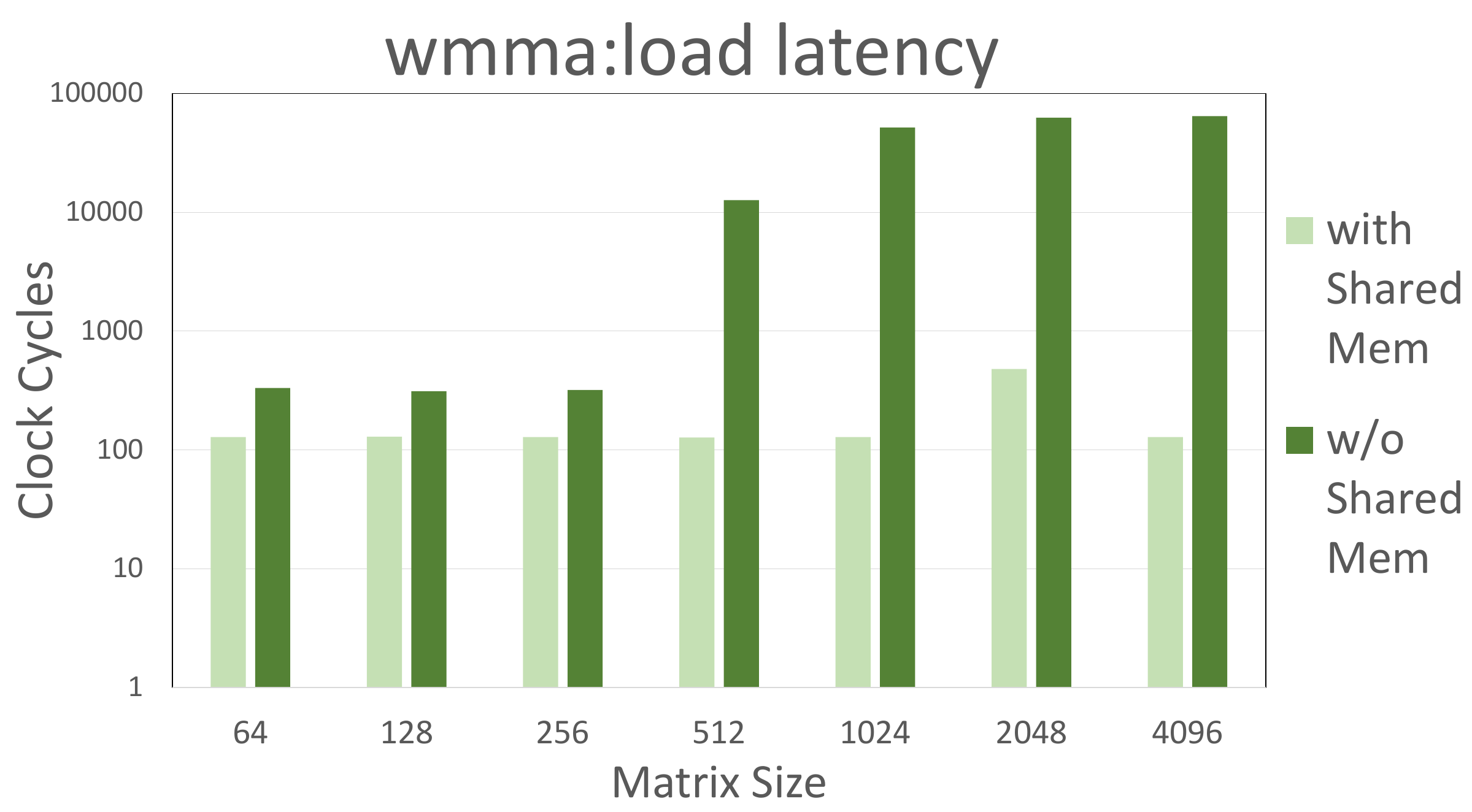}
\end{subfigure}\hspace{0.1in}
\begin{subfigure}{.3\textwidth}
\centering\includegraphics[width=1\textwidth]{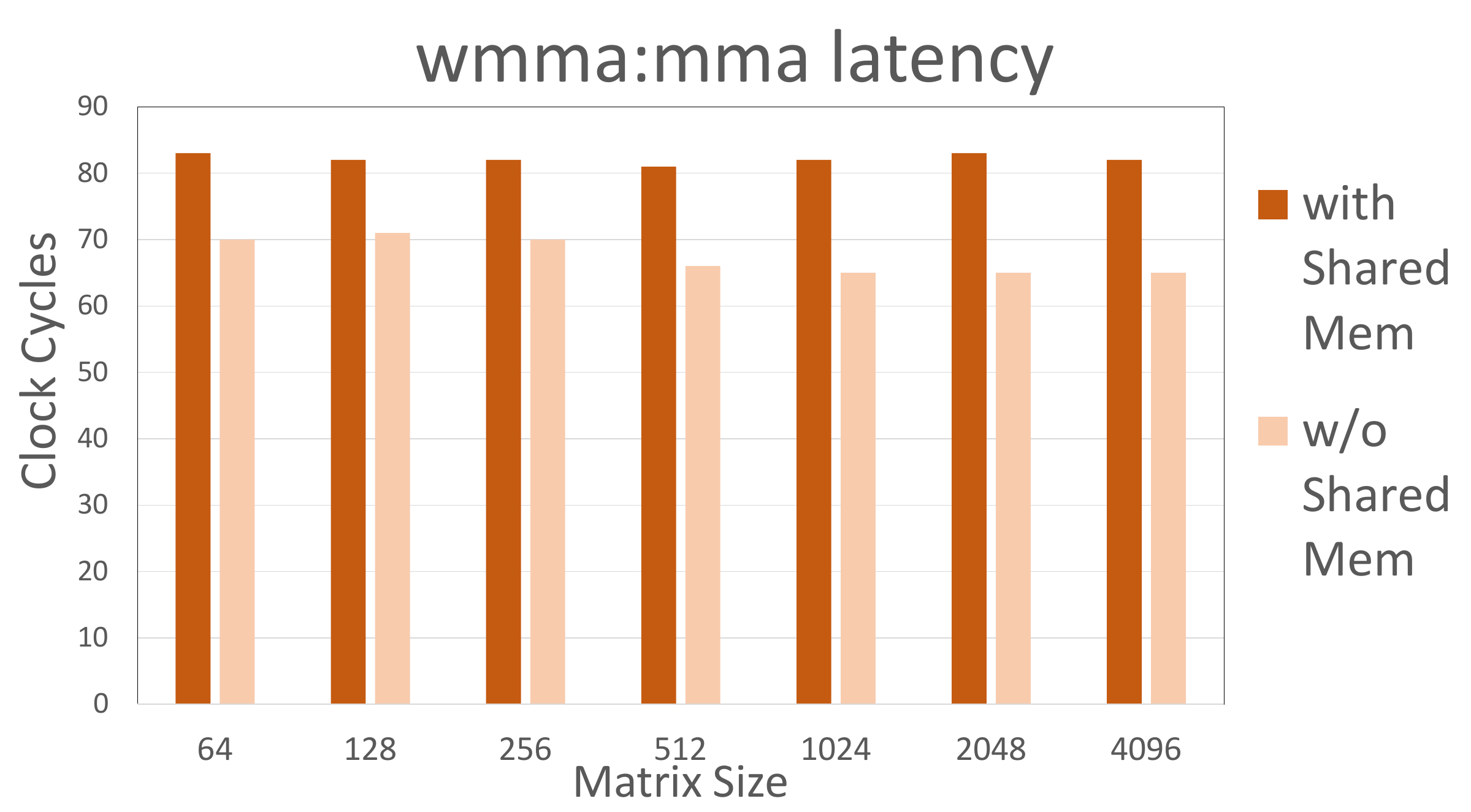}
\end{subfigure}\hspace{0.1in}
\begin{subfigure}{.3\textwidth}
\centering\includegraphics[width=1\textwidth]{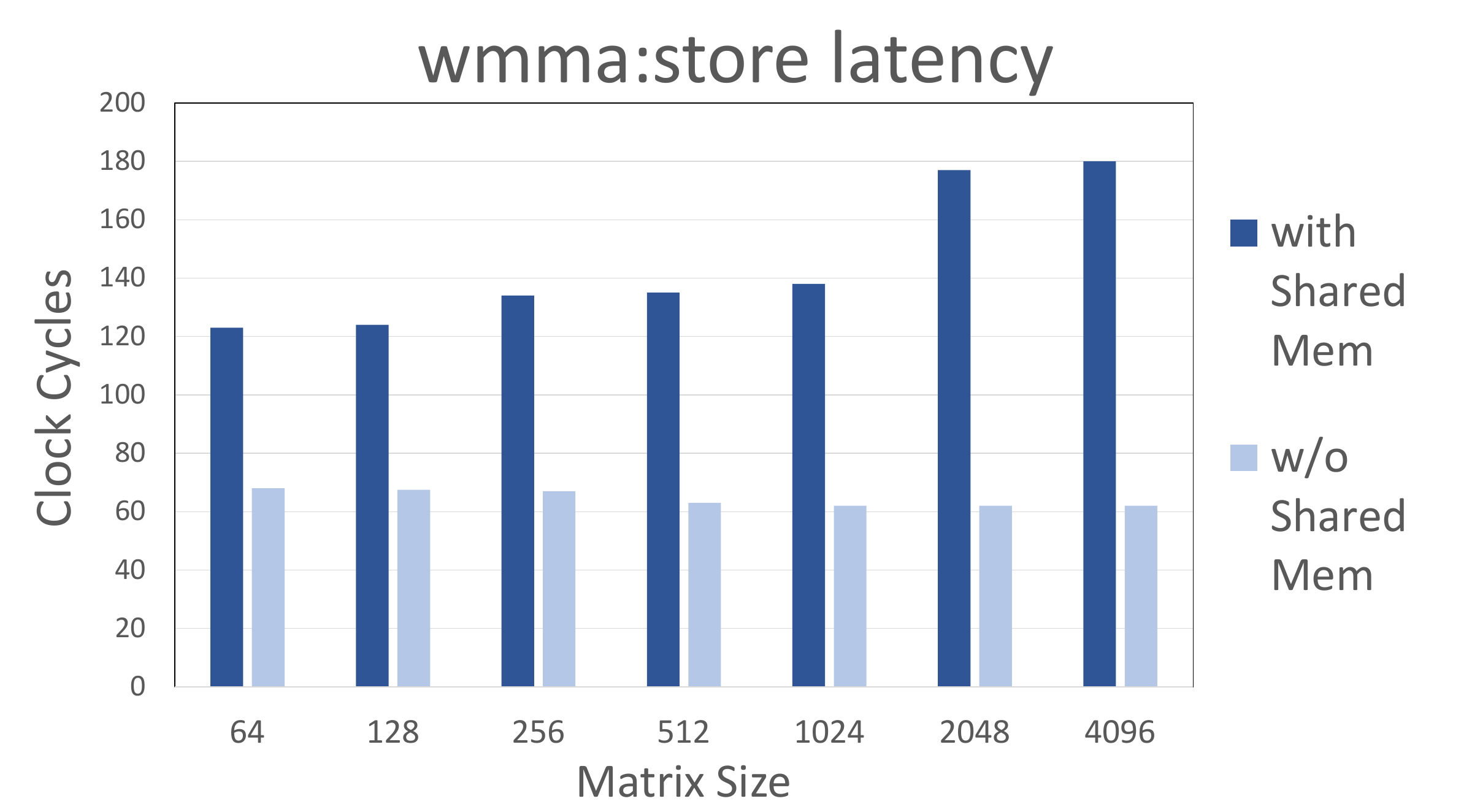}
\end{subfigure}\hspace{0.1in}
\caption{Variation of latency of \lstinline{wmma.load}, \lstinline{wmma.mma} and \lstinline{wmma.store} with matrix size}
\label{latencyofwmmaptx}
\end{figure*}

Our changes to model the tensor cores in Volta are available in the ``dev''
branch of GPGPU-Sim~\cite{bakhoda2009analyzing} on
github\footnote{\url{https://github.com/gpgpu-sim/gpgpu-sim_distribution/tree/dev}}.
We extended the current version of GPGPU-Sim to support 16-bit floating-point
by using a half-precision C++ header-only library~\cite{HalfPrecisionLibrary}. 
The library provides an efficient
implementation of 16-bit floating-point conforming to the IEEE 754 half-precision format.
It provides common arithmetic operations and type conversion. 
GPGPU-Sim currently only supports SASS execution for the G90 architecture; therefore, 
we only model tensor core operations at the PTX level.  To do so, we added functional
and timing models for the \lstinline{wmma.load}, \lstinline{wmma.mma} and \lstinline{wmma.store}
PTX instructions described in Section~\ref{sec:ptx}.

Our functional model of the \lstinline{wmma.load} and \lstinline{wmma.store} PTX instructions support all
possible layout combinations for operand matrix A, B and C.  Our functional
model follows the operand matrix element to thread mapping shown in Figure~\ref{Mapping}.
We have verified the timing model generates the exact same number of coalesced memory
transactions generated by the Titan~V GPU for these operations.

Our functional model of the \lstinline{wmma.mma} instruction supports all 32 possible configurations
supported on the Titan~V GPU. 
A timing model for the tensor core functional unit is added to the GPU pipeline.
We interface our tensor core timing model to the operand collector unit modeled in GPGPU-Sim. 
Each \lstinline{wmma.mma} instruction is issued to the tensor core unit after all of its source
operands are ready in the operand collector. 
We updated the scoreboard to check for RAW and WAW hazard associated with~\lstinline{wmma.mma} instructions.

We validate our tensor core model by comparing against an NVIDIA Tesla Titan V with
CUDA Capability 7.0, hosted by an Intel Core i7-4771 3.50GHz based workstation with
Ubuntu 16.04.4 LTS, CUDA Toolkit Version 9.0, NVIDIA 410.48 GPU driver, and gcc 4.9.4. 
Figure~\ref{modelling} compares the cycles required to execute a WMMA based matrix-multiply and accumulate kernel 
on the Titan V GPU and GPGPU-Sim as matrix size varies.  We find 
GPGPU-Sim tracks real hardware very accurately with a standard deviation of less than 5\%. This is
despite the fact our model is implemented at the PTX level.

\begin{figure*}
\centering
\centering\includegraphics[width=0.7\textwidth]{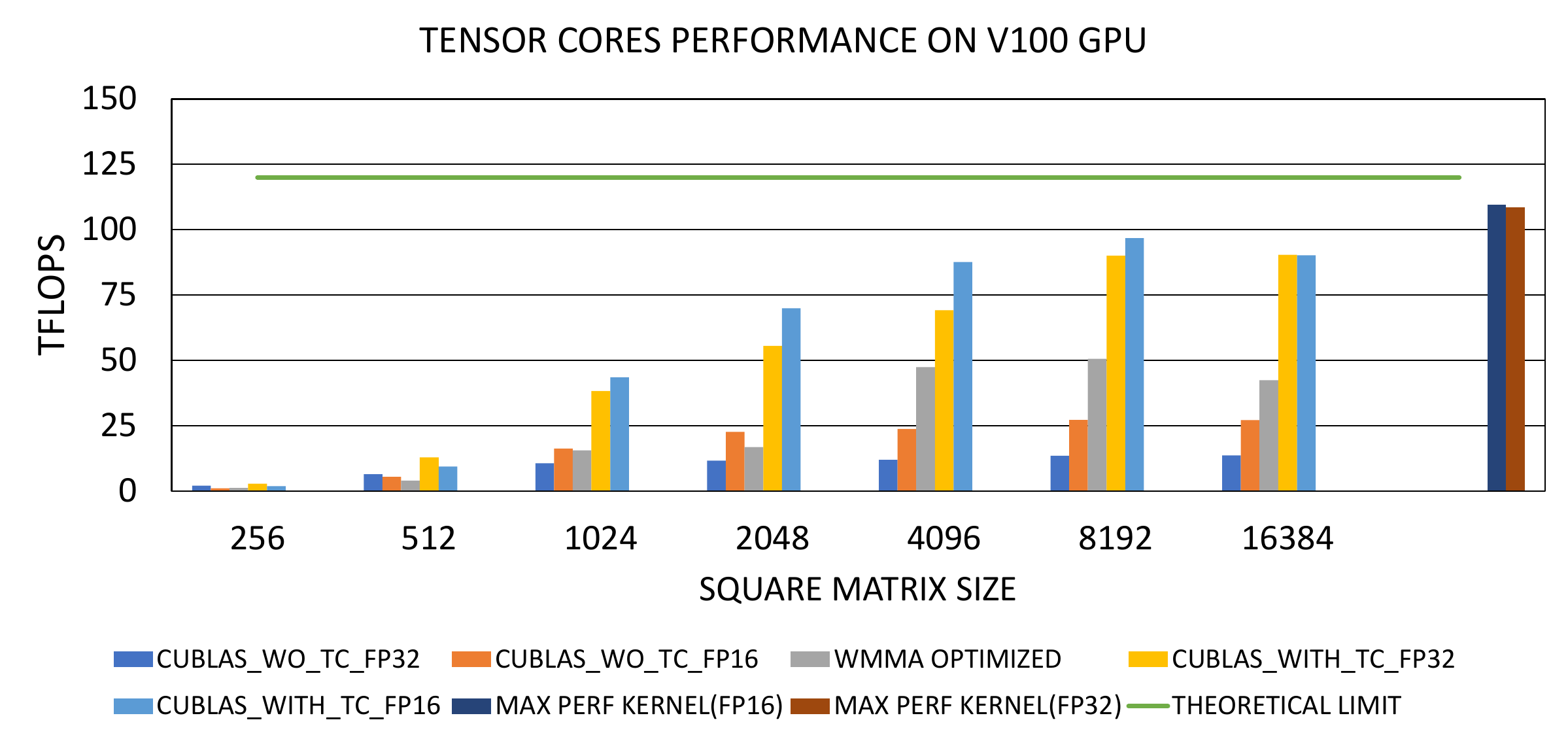}
\caption{Tensor Cores Performance}
\label{tensorcoreperformance}
\end{figure*}

\subsection{CUTLASS} 

CUTLASS is an open-source CUDA C++ template
library for efficient linear algebra in C++.
It provides basic building block for implementing high-performance fused matrix-multiply kernels
for deep learning.

We modified GPGPU-Sim to enable it to run CUTLASS including adding missing API calls and PTX
instruction definitions.
NVIDIA developed a unit-test suite for CUTLASS library consisting
of around 680 test cases. We verified these test cases run with our modifications to
GPGPU-Sim\footnote{\url{https://github.com/gpgpu-sim/cutlass-gpgpu-sim}}.
Figure~\ref{cutlass:a} shows a comparison of
Instructions Per Cycle (IPC) measured 
on GPGPU-Sim versus a real NVIDIA Titan~V GPU 
for a tensor core enabled kernel developed using CUTLASS.
This data shows an IPC correlation of 99.60\%.
Figure~\ref{cutlass:b} shows GPGPU-Sim tends to have higher
performance versus hardware as matrix size increases.

\subsection{Profiling Tensor Cores}

In this section we measure the performance gain 
obtained when employing tensor cores measured on a real NVIDIA Titan~V GPU.

NVIDIA's documentation suggests that tensor cores can provide peak theoretical
performance of 125 TFLOPs.  The maximum performance we obtained for a GEMM kernel 
was around 96 TFLOPs. This performance was observed
for $8192\times 8192$ matrix using FP16 mode. 
To measure the maximum sustainable tensor core throughput we developed a
kernel with repeated \lstinline{wmma.mma} operations 
(computational intensity on the order of $10^{8}$). 
The performance obtained 
was 109.6 TFLOPs in FP16 mode and 108.7 TFLOPs in mixed-precision mode.

Figure~\ref{latencydist} shows the results of profiling the 
latency of \lstinline{wmma.load}, \lstinline{wmma.mma} and \lstinline{wmma.store} instructions
during several iterations of a WMMA kernel.  This kernel uses shared memory and performs matrix-multiply accumulate
operations on a $1024\times1024$ matrix.  All three graphs show occasional high latencies.  These
may result from some combination of warp scheduling policies and high memory traffic. 
We find the minimum latency of \lstinline{wmma.load}, \lstinline{wmma.store} and \lstinline{wmma.mma} instructions
is 125, 120 and 70 clock cycles respectively.

In Figure~\ref{latencyofwmmaptx} we plot the median latency 
to analyze how \lstinline{wmma.load}, \lstinline{wmma.mma} and \lstinline{wmma.store}
latency varies with the matrix size for WMMA kernels. The \lstinline{wmma.load}
latency is plotted with a logarithmic axis.  
Using shared-memory reduces median \lstinline{wmma.load} latency by more
than 100$\times$ when operating on a larger matrix.

Figure~\ref{tensorcoreperformance} shows the performance achieved by the tensor
cores in different scenarios:
In this figure we compare performance
of a GEMM kernel implemented with various APIs (CUBLAS, WMMA) with (WITH) or without (WO) tensor
cores (TC) using mixed-precision (FP32) or FP16 mode.  In this graph
``MAX PERF KERNEL'' is our kernel designed to stress tensor core performance in FP16 or mixed-precision (FP32) mode.
THEORETICAL LIMIT is the peak performance of 125 TFLOPs.
The WMMA GEMM includes optimizations like
using shared memory and proper memory layout. The performance gain obtained
using the cuBLAS GEMM kernel is more than the WMMA GEMM implementation (both the
kernels using tensor cores). cuBLAS is a highly optimized library which
has optimizations to avoid shared memory bank conflicts and employs
software pipelining. We find tensor cores provide a performance boost of about
$3-6\times$ times that of SGEMM (Single Precision GEMM) kernel and about
$3\times$ that of HGEMM (Half Precision GEMM).



\section{Related Work} 

This section briefly discusses related work.
Wong et al.~\cite{wong2010demystifying} performed a thorough analysis of the NVIDIA GT200
using an extensive set of microbenchmarks.  They explored architectural details of the processing cores and the memory 
hierarchies.
The describe previously undisclosed details of barrier synchronization and the memory hierarchy including 
TLB organization in GPUs.
Jia et al.~\cite{jia2018dissecting} explored tensor cores in detail.
They decoded sets and steps for Volta tensor cores in mixed-precision mode.
In contrast, we comprehensively investigated both modes of operation. We found that
sets and steps behave differently in FP16 mode than in mixed precision mode.
We uncovered the organization of theadgroups into octets.
We determined the mapping of operand matrix elements to threads for the tensor cores in the Turing architecture
and found they behaves differently from the Volta tensor cores.  
We also provide a methodology for uncovering the information presented (including describing our microbenchmarks).
Markidis et al.~\cite{markidis2018nvidia} studied the impact of precision loss and programmability aspect of Tensor Cores 
for HPC application.
Khairy, et al.~\cite{MemSystemDesign} studied the memory system of modern GPUs including Volta 
and discovered many important design decisions in the memory system. They modeled it in GPGPU-Sim and 
achieve a very high correlation on a wide range of GPGPU workloads.

\section{Conclusion} 

In this paper we investigated the design of the tensor core machine learning
accelerators integrated into recent GPUs from NVIDIA.  We performed a 
detailed characterization and analysis of the tensor cores implemented in NVIDIA's
Volta and Turing architectures.  
This analysis guided the development of a detailed architectural model.
We implemented a model for the Volta tensor cores in GPGPU-Sim and found
its performance agreed well with hardware, obtaining a 99.6\% IPC correlation
versus a Titan V GPU.  As part of our efforts we also enabled CUTLASS,
NVIDIA's open-source CUDA C++ template library supporting tensor cores, on GPGPU-Sim. 
We believe that combined the above work will serve as a promising starting point 
for further micro-architectural investigation of machine learning workloads.

\section*{Acknowledgment}
We thank Francois Demoullin, Deval Shah, Dave Evans, Bharadwaj Machiraju, Yash Ukidave and the anonymous reviewers for their valuable comments on this work. This research has been funded in part by the Computing Hardware for Emerging Intelligent Sensory Applications (COHESA) project. COHESA is financed under the National Sciences and Engineering Research Council of Canada (NSERC) Strategic Networks grant number NETGP485577-15.

\bibliographystyle{ieeetr}
\bibliography{citation.bib}

\end{document}